\documentclass[11pt,letterpaper]{article}

\usepackage{amsfonts}
\usepackage{amsmath}
\usepackage{amssymb}
\usepackage{enumerate}
\usepackage{natbib}
\usepackage{color}
\usepackage{graphicx,subfigure}
\usepackage{epstopdf}
\usepackage{mathtools}                
\mathtoolsset{showonlyrefs=true}
\usepackage{fixltx2e,amsmath}                                           

\MakeRobust{\eqref}
\usepackage{amssymb}                            
\usepackage[mathscr]{eucal}                     
\usepackage{setspace}
\usepackage{hyperref}
\newcommand\E{\ensuremath{\mathbb{E}}}
\newcommand\R{\ensuremath{\mathbb{R}}}

\newcommand\F{\mathcal{F}}

\newcommand\half{\frac{1}{2}}
\newcommand{\dx}[1]{ \,d#1 }

\newcommand{\indic}[1]{1\hspace{-2.1mm}{1}_{\{#1\}}} 

\DeclareMathOperator*{\argmax}{arg\,max}

\def\R{{\mathbb R}}

\def\P{{\mathbb P}}

\def\1{{\mathbf 1}}
\def\F{{\mathcal F}}

\def\L{{\mathcal L}}
\def\W{{\mathcal W}}

\def\setT{{\mathcal T}}
\newtheorem{theorem}{Theorem}

\newtheorem{lemma}[theorem]{Lemma}

\newtheorem{remark}[theorem]{Remark}

\numberwithin{equation}{section}
\numberwithin{theorem}{section}
\begin{document}
\title{Optimal Multiple Trading Times Under the Exponential OU Model with  Transaction Costs\thanks{We thank  two anonymous referees for their thorough reviews of our paper and their helpful remarks. }}
\author{Tim Leung\thanks{IEOR Department, Columbia University, New York, NY 10027; email:\,\mbox{tl2497@columbia.edu}. } \and Xin Li\thanks{IEOR Department, Columbia University, New York, NY 10027; email:\,\mbox{xl2206@columbia.edu}.} \and Zheng Wang\thanks{IEOR Department, Columbia University, New York, NY 10027; email:\,\mbox{zw2192@columbia.edu}.}}
\date{\today}
\maketitle
\begin{abstract}
This paper studies the timing of trades under mean-reverting price dynamics subject to fixed transaction costs. We solve an optimal double stopping problem to determine the optimal times to enter and subsequently exit the market, when prices are driven by an exponential Ornstein-Uhlenbeck   process. In addition, we  analyze a related optimal switching problem that involves  an infinite sequence of trades, and identify the conditions  under which the  double stopping and switching problems admit the same optimal entry and/or exit timing strategies.    Among our results, we find that   the investor generally enters when the price is low, but may find it  optimal to wait if the current price  is sufficiently  close to zero. In other words, the continuation (waiting) region for entry is \emph{disconnected}.   Numerical results are provided to illustrate the dependence of timing strategies on  model parameters and transaction costs. 
\end{abstract}\vspace{10pt}

\begin{small}
\end{small}


\section{Introduction}
One important problem commonly faced by  individual and institutional investors  is to determine when to buy and sell an asset.  As  a potential investor observes the price process of an asset, she can decide to enter the market immediately or wait for a future opportunity. After completing the first trade, the investor will need  to select  the time to close  the position. This motivates  us to investigate the optimal sequential timing of trades.

Naturally, the optimal sequence of trading times should depend on the price dynamics of the risky asset. For instance, if the price process is a super/sub-martingale, then the investor, who seeks to maximize the expected liquidation value,  will  either sell immediately or wait forever. Such a trivial timing arises when the underlying price follows a geometric Brownian motion. Similar observations can also be found in,  among others, \cite{shiryaev2008thou}. On the other hand, it has been widely observed that  many asset prices exhibit mean reversion, ranging from  equities and commodities to currencies and volatility indices  (see \cite{metcalf1995investment}, \cite{bessembinder1995}, \cite{Casassus2005}, and references therein). To   incorporate mean-reversion for positive price processes, one popular choice for pricing and investment applications  is the exponential Ornstein-Uhlenbeck (XOU) model, as proposed by \cite{schwartz1997stochastic} for commodity prices,   due to its analytical  tractability. It also serves as the building block of more sophisticated mean-reverting models.
  
  In this paper, we study  the optimal timing of trades under  the XOU model subject to transaction costs.  We consider two different but related  formulations. First, we consider the trading problem with a single entry and single exit with a fixed cost incurred at each transaction.  This leads us to analyze an   \emph{optimal double stopping} problem. In the second formulation, the investor is assumed to enter and exit the market infinitely many times with transaction costs. This gives rise to an \textit{optimal switching} problem.

We analytically derive the  non-trivial entry and exit timing strategies.  Under both approaches, it is optimal to sell when the asset  price is sufficiently high, though at different levels.  As for entry timing,  we find that, under some  conditions, it is optimal for the investor not to enter the market at all when facing the  optimal switching problem. In this case for  the investor who has a long position,  the optimal switching problem reduces  into an optimal stopping problem, where  the optimal liquidation level is identical to that of  the  optimal double stopping problem. Otherwise, the optimal entry timing strategies for the double stopping and  switching problem are  described by the underlying's first passage time to an interval that lies above level zero. In other words,  the continuation region for entry is \emph{disconnected} of the form $(0, A)\cup(B, +\infty)$, with critical price levels $A$ and $B$ (see Theorems \ref{thm:optEntryexpOU} and \ref{thm:XOU2} below).  This means that  the investor generally enters when the price is low, but may find it  optimal to wait if the current price is too close to zero. We find that this phenomenon is a distinct consequence  due to fixed transaction costs under the XOU model. Indeed, when there is no fixed costs, even if there are  proportional transaction costs (see \cite{zhang2008trading}), the entry timing is simply characterized by a single  price level. 

A typical solution  approach for optimal stopping problems driven by diffusion involves the analytical and numerical  studies of  the associated free boundary problems or variational inequalities (VIs); see e.g.  \cite{Bensoussan}, \cite{Oksendal2003}, and \cite{sun1992nested}.   This approach is very useful when the structure of the optimal buy/sell strategies are known.  In contrast to the VI approach,  we solve  the double stopping problem by characterizing the  value functions (for entry and exit)    as the smallest concave majorant of the corresponding   reward functions (see \cite{dayanik2003optimal} and references therein).  This allows us to directly construct the value function, without \emph{a priori} finding a candidate value function or imposing conditions on the stopping and continuation regions, such as whether they are connected or not.  In other words, our method will derive  the structure of the stopping and continuation regions as an output. Moreover,  the VI method becomes  more challenging when the form of the reward function does not possess some amenable properties, such as convexity, monotonicity, and positivity. This gives another reason for our probabilistic approach since the reward function for the entry problem is neither convex/concave, nor  monotone, nor always positive.  Having solved the optimal double stopping problem, we determine the optimal structures  of the buy/sell/wait regions. We then apply this to infer a similar solution structure for the optimal switching problem and verify using the variational inequalities.

 In the literature, \cite{zhang2008trading}
investigate the optimal switching problem under the XOU price dynamics,  with slippage  (proportional transaction cost). As extension, \cite{kong2010optimal}  allow for short selling so that the investor can enter the market with either long or short position, and close it out in the next trade. As a numerical approach, \cite{song2009stochastic} discuss  a stochastic approximation  scheme to compute  the optimal buying and selling price levels by a priori assuming a buy-low-sell-high strategy under the XOU model. In contrast to these studies, we study both optimal double stopping and switching problems specifically under exponential OU with  fixed transaction costs. In particular, the optimal entry timing with fixed transaction costs   is characteristically different from that with slippage.

\cite{zervos2011buy} consider an optimal switching problem with fixed transaction costs  under  a class of time-homogeneous diffusions, including the GBM,  mean-reverting CEV  underlying, and other models. However, their results are not applicable to the exponential OU model  as it violates Assumption 4 of their paper (see also Remark \ref{XOUZervos} below).  Indeed, their model assumptions restrict    the optimal entry  region to be  represented  by a single critical threshold, whereas we show that in the XOU model the optimal entry  region is characterized by two positive price levels. 


As for related applications of optimal double stopping,   \cite{LeungLi2014OU} study the optimal timing to trade an OU price spread with a stop-loss constraint. \cite{KS2007} analyze  the double stopping times for a  risk process from the insurance company's perspective.  The problem of  timing to buy/sell derivatives has also been studied in \cite{LeungLudkovski2011} (European and American options) and \cite{leungLiu} (credit derivatives). \cite{menaldi1996optimal} study an optimal starting-stopping problem for general Markov processes, and provide the mathematical characterization of the value functions.

The rest of the paper is structured as follows. In Section \ref{sect-overview}, we formulate both the optimal double stopping and optimal switching problems. Then, we present our analytical and numerical results  in Section \ref{sect-solution}. The   proofs of our main  results are detailed in Section \ref{sect-method}. Finally, the Appendix contains the proofs for   a number of lemmas.

\section{Problem Overview}\label{sect-overview}
In the background, we fix a probability space $(\Omega, \F, \P)$, where $\P$ is   the historical probability measure. In this section, we provide an overview of our optimal double stopping and switching problems, which will involve an exponential Ornstein-Uhlenbeck (XOU) process. The XOU process $(\xi_t)_{t \geq 0}$ is defined by 
\begin{align}\label{XOU} 
\xi_t = e^{X_t}, \qquad dX_{t}=  \mu( \theta - X_t)\,dt+\sigma \,dB_{t}.
\end{align}
Here, $X$ is an OU process driven by a standard Brownian motion $B$, with  constant parameters $\mu, \sigma>0$, $\theta \in \R$. In other words,  $X$ is the log-price of the positive XOU process $\xi$.


\subsection{Optimal Double Stopping Problems}
   Given a risky asset with an XOU price process, we first consider the optimal timing to sell.  If the share of the asset is sold at some time $\tau$, then the investor will receive the  value  $\xi_\tau = e^{X_\tau}$  and pay a constant transaction cost $c_s > 0$. Denote by $\mathbb{F}$ the filtration generated by $B$, and $\setT$    the set of all $\mathbb{F}$-stopping times.  To maximize the expected discounted value,  the investor solves the optimal stopping problem
\begin{align}V(x) = \sup_{\tau \in \setT}\E_x\!\left\{e^{-r \tau}(e^{X_{\tau}} - c_s) \right\}, \label{V1a}\end{align}
where   $r>0$ is the constant discount rate, and  $\E_x\{\cdot\}\equiv\E\{\cdot|X_0=x\}$.

The value function $V(x)$ represents the expected  liquidation
value associated with  $\xi$. On the other hand, the current price plus the
transaction cost constitute  the total cost to enter the trade. Before even holding the risky asset, the investor
can always choose the optimal timing to start the trade, or not to enter at
all. This leads us to analyze the  entry timing inherent in the trading
problem. Precisely, we solve \begin{align}J(x) =  \sup_{\nu \in \setT
}\E_x\!\left\{e^{-r \nu} (  V(X_{\nu})  - e^{X_{\nu}}- c_b)\right\},
\label{J1a}\end{align} with the constant transaction cost $ c_b> 0$ incurred  at the time of purchase.  In other words, the trader seeks
to maximize the expected difference between the value function
$V(X_\nu)$ and the current  $e^{X_{\nu}}$, minus transaction cost  $c_b$. The value function $J(x)$
represents the maximum expected value of the investment opportunity in
the price  process $\xi$, with transaction costs $c_b$ and $c_s$ incurred, respectively, at entry and exit. For  our analysis, the transaction costs $c_b$ and $c_s$ can be different. To facilitate presentation, we denote the functions
\begin{align} \label{hXOU1}h_s(x) = e^x -c_s \quad \text{ and } \quad h_b(x)= e^x + c_b.\end{align}


If it turns out that $J(X_0)\le  0$  for some initial value
$X_0$, then the investor will not start to trade $X$. It is important to identify the trivial cases under any given dynamics. Under the XOU model, since   $\sup_{x\in\R}( V(x) -h_b(x))\leq 0$
implies that   $J(x) \leq 0$ for $x \in \R$, we shall therefore focus on the  case with
\begin{align}\label{generalassume} \sup_{x\in\R}( V(x)  -h_b(x))> 0,
\end{align} and solve for the non-trivial optimal  timing strategy.

\subsection{Optimal Switching Problems}\label{ISF}
Under the optimal switching approach,  the investor is assumed to commit to an infinite number of trades. The sequential trading times are modeled  by the stopping times  $\nu_1,\tau_1,\nu_2,\tau_2,\dots \in \setT$ such that
\begin{align*}
0\leq \nu_1 \leq \tau_1 \leq \nu_2 \leq \tau_2 \leq \dots.
\end{align*}
A share of the risky asset is bought and sold, respectively, at times  $\nu_i$ and   $\tau_i$, $i\in\mathbb{N}$. The investor's optimal timing to trade would depend  on the  initial position. Precisely, under the XOU model,  if the investor starts with a zero    position, then the first trading decision is when to \emph{buy} and  the corresponding optimal switching problem is
\begin{align}\label{J}
\tilde{J}(x) = \sup_{\Lambda_0}\E_x\left\{\sum_{n=1}^\infty [e^{-r\tau_n}h_s(X_{\tau_n}) - e^{-r \nu_n} h_b(X_{\nu_n})] \right\},
\end{align}
with   the set of admissible stopping times $\Lambda_0=(\nu_1,\tau_1,\nu_2,\tau_2,\dots)$, and  the reward functions $h_s$ and $h_b$   defined in \eqref{hXOU1}.  On the other hand, if the investor is initially holding a share of the asset, then the investor first determines when to \emph{sell} and solves
\begin{align}\label{V}
\tilde{V}(x) = \sup_{\Lambda_1}\E_x\left\{e^{-r\tau_1}h_s(X_{\tau_1}) + \sum_{n=2}^\infty [e^{-r\tau_n}h_s(X_{\tau_n}) - e^{-r \nu_n} h_b(X_{\nu_n})] \right\},
\end{align}
with  $\Lambda_1=(\tau_1,\nu_2,\tau_2,\nu_3,\dots)$.
%


 In summary, the optimal  double stopping and switching problems differ in the number of trades.  Observe that any strategy for the double stopping problems \eqref{V1a} and \eqref{J1a} are also candidate strategies for the switching problems \eqref{V} and \eqref{J} respectively. Therefore, it follows that  $V(x) \leq \tilde{V}(x)$ and  $J(x) \leq \tilde{J}(x).$ Our objective is to derive and compare the corresponding optimal timing strategies under these two approaches.  

\section{Summary of Analytical Results}\label{sect-solution}
We first summarize our analytical results and illustrate the optimal trading strategies. The  method of solutions and their proofs will be discussed in Section \ref{sect-method}.
We begin with the optimal stopping problems \eqref{V1a} and \eqref{J1a} under the XOU model. Denote  the infinitesimal generator of the OU process $X$ in \eqref{XOU} by
\begin{align}\label{genX}\L = \frac{\sigma^2}{2}\frac{d^2}{d x^2} + \mu(\theta - x)\frac{d}{d x}.
\end{align}
Recall that the classical  solutions of the differential equation \begin{align}\L u(x)=ru(x),\label{LUXOU}\end{align} for $x\in \R$, are (see e.g.  p.542 of \cite{borodin2002handbook} and Prop. 2.1 of \cite{alili2005representations}):
\begin{align}
F(x) \equiv F(x;r):= \int_0^\infty u^{\frac{r}{\mu}-1} e^{\sqrt{\frac{2\mu}{\sigma^2}}(x-\theta)u-\frac{u^2}{2}} \dx{u},\label{FOU}\\
G(x) \equiv G(x;r):= \int_0^\infty u^{\frac{r}{\mu}-1} e^{\sqrt{\frac{2\mu}{\sigma^2}}(\theta-x)u-\frac{u^2}{2}} \dx{u}.\label{GOU}\end{align}
Direct differentiation yields that   $F'(x) >0$,  $F''(x)>0$,  $G'(x)<0$ and $G''(x)>0$. Hence, we observe that both $F(x)$ and $G(x)$ are strictly positive and convex, and they are, respectively, strictly increasing and decreasing.


Define the first passage time of $X$ to some level $\kappa$ by $\tau_{\kappa} = \inf\{t \geq 0: X_t = \kappa\}$. As is well known, $F$ and $G$ admit the probabilistic expressions (see \cite{ito1965diffusion} and \cite{Rogers2000}):
\begin{align}\label{generalFG1}
\E_x\{e^{-r\tau_\kappa}\} = \begin{cases}
\frac{F(x)}{F(\kappa)} &\, \textrm{ if }\, x\leq \kappa,\\
\frac{G(x)}{G(\kappa)} &\, \textrm{ if }\, x\geq\kappa.
\end{cases}
\end{align}

%

\subsection{Optimal Double Stopping Problems}\label{sect-doubleXOU}
We now present the result for the optimal exit timing problem.
\begin{theorem}\label{thm:optLiquexpOU} The optimal liquidation problem \eqref{V1a} admits the solution
\begin{align}V(x) =
\begin{cases}
 \frac{e^{b^{*}}-c_s}{F(b^{*})}F(x) &\, \textrm{ if }\, x < b^{*},\\
e^x-c_s &\, \textrm{ if }\, x \geq b^{*},
\end{cases}\label{VexpOUsol}\end{align}
where the optimal log-price level $b^{*}$ for liquidation is uniquely found from the equation 
\begin{align}\label{eq:solvebexpOU}
e^{b}F(b)=(e^{b}-c_s)F'(b).
\end{align}
The optimal liquidation time is given by 
\begin{align}\tau^{*}&= \inf \{ \, t\ge 0 \,:\, X_t   \ge b^{*}\,\}= \inf \{ \,t\ge 0 \,:\,\xi_t   \ge e^{b^{*}}\,\}. \label{taustar1}
\end{align}
\end{theorem}



%

\newpage 

\begin{theorem}\label{thm:optEntryexpOU}
Under the XOU model, the optimal entry timing
problem \eqref{J1a} admits the solution
\begin{align}\label{JsoluexpOU}
  J(x) =\displaystyle
\begin{cases}
  PF(x) &\, \textrm{ if }\, x \in (-\infty,a^{*}),\\
  V(x)-(e^x+c_b) &\, \textrm{ if }\, x \in [a^{*},d^{*}],\\
  QG(x) &\, \textrm{ if }\, x \in (d^{*},+\infty),
\end{cases}
\end{align}
with the constants
\begin{align}
P = \frac{V(a^{*})-(e^{a^{*}}+c_b)}{F(a^{*})}, \quad
Q &=
\frac{V(d^{*})-(e^{d^{*}}+c_b)}{G(d^{*})},
 \end{align}
 and the   critical levels $a^{*}$ and $d^{*}$ satisfying, respectively,
\begin{align}\label{eq:solveaexpOU}
 F(a)({V}^\prime\!(a)  - e^a) = F'(a)(V(a)-(e^a+c_b)),\\
 G(d)({V}^\prime\!(d)  - e^d) = G'(d)(V(d)-(e^d+c_b)).\label{eq:solvedexpOU}
\end{align}
The optimal entry time is given by
\begin{align}\nu_{a^{*},d^{*}} := \inf \{  t\ge 0 \,:\, X_t \in [a^{*}, d^{*}]\}.\label{taustar2}\end{align}
\end{theorem}

In summary, the investor should  exit the market as soon as the price reaches the upper level $e^{b^*}$. In contrast, the optimal entry timing is the first time that the XOU price $\xi$ enters the interval $[e^{a^{*}},e^{d^{*}}]$. In other words, it is optimal to  wait if the current price $\xi_t$ is too close to zero, i.e. if $\xi_t < e^{a^{*}}$. Moreover, the  interval $[e^{a^{*}},e^{d^{*}}]$ is contained in   $(0, e^{b^{*}})$, and thus, the continuation  region for market entry is \emph{disconnected}.

\subsection{Optimal Switching Problems}\label{sect-switchingXOU}
We now turn to the optimal switching problems defined in \eqref{J} and \eqref{V} under the XOU model. To facilitate  the presentation, we denote
\begin{align}
f_s(x) := (\mu\theta+\half \sigma^2-r)-\mu x+r c_s e^{-x}, \label{fs}\\
f_b(x) := (\mu\theta+\half \sigma^2-r)-\mu x-r c_b e^{-x}. \label{fb}
\end{align}
Applying the operator $\mathcal{L}$ (see \eqref{genX}) to $h_s$ and $h_b$ (see \eqref{hXOU1}), it follows  that $(\L-r)h_s(x) = e^xf_s(x)$ and $(\L-r)h_b(x) = e^xf_b(x)$. Therefore, $f_s$ (resp. $f_b$) preserves the sign of $(\L-r)h_s$ (resp. $(\L-r)h_b$).  It can be shown that $f_s(x)=0$ has a unique root, denoted by $x_s$. However, 
\begin{equation}\label{eqn:XOUfb}
f_b(x)=0
\end{equation}
may have no root, a single root, or two distinct roots, denoted by $x_{b1}$ and $x_{b2}$, if they exist. The following observations will also be useful:
\begin{align}\label{Lr}
f_s(x) \begin{cases}
>0 &\, \textrm{ if }\, x<x_s,\\
<0 &\, \textrm{ if }\, x>x_s,
\end{cases} \quad \textrm{and} \quad
f_b(x) \begin{cases}
<0 &\, \textrm{ if }\, x\in (-\infty, x_{b1}) \cup (x_{b2}, +\infty),\\
>0 &\, \textrm{ if }\, x\in (x_{b1}, x_{b2}).
\end{cases}
\end{align}


The optimal switching problems have two different sets of solutions depending on the problem data.  

\begin{theorem}\label{thm:XOU1}
The optimal switching problems \eqref{J} and \eqref{V} admit the solutions
\begin{align}\label{sol:XOU}
\tilde{J}(x) = 0, \, \text{ for }  x \in \R,  \quad \text{ and  } \quad 
\tilde{V}(x) =\displaystyle
\begin{cases}
\frac{e^{b^*}-c_s}{F(b^*)} F(x) &\, \textrm{ if }\, x < b^*,\\
e^x-c_s &\, \textrm{ if }\, x \geq b^*,
\end{cases}
\end{align}
where $b^*$ satisfies \eqref{eq:solvebexpOU}, if any of the following mutually exclusive conditions holds:
\begin{enumerate}[(i)]
\item \label{XOU1} There is no root or a single root to equation \eqref{eqn:XOUfb}.

\item \label{XOU3}  There are two distinct roots to \eqref{eqn:XOUfb}. Also
\begin{align}\label{eq:XOUa}
\exists \, \tilde{a}^* \in (x_{b1}, x_{b2}) \quad \textrm{ such that } \quad F(\tilde{a}^*)e^{\tilde{a}^*} = F'(\tilde{a}^*)(e^{\tilde{a}^*}+c_b),
\end{align}
and
\begin{align}\label{XOUcond3}
\frac{e^{\tilde{a}^*}+c_b}{F(\tilde{a}^*)} \geq \frac{e^{b^*} - c_s}{F(b^*)}.
\end{align}

\item \label{XOU2} There are two distinct roots to \eqref{eqn:XOUfb} but \eqref{eq:XOUa} does not hold.

\end{enumerate}
\end{theorem}

In Theorem \ref{thm:XOU1}, $\tilde{J} = 0$ means  that it is  optimal not to enter the market at all. On  the other hand, if one starts with a unit  of the underlying asset, the optimal switching problem reduces to a problem of optimal single stopping. Indeed, the investor will never re-enter the market after exit.   This is identical to the optimal liquidation problem \eqref{V1a} where there is only a single (exit) trade. The optimal strategy in this case is the same as $V$ in \eqref{VexpOUsol} -- it is optimal to exit the market as soon as the log-price $X$ reaches the threshold  ${b^*}$.

We also address the remaining case when none of the conditions in Theorem \ref{thm:XOU1} hold. As we show next, the optimal strategy will involve both entry and exit thresholds.
\begin{theorem}\label{thm:XOU2}
If there are two distinct roots to \eqref{eqn:XOUfb}, $x_{b1}$ and $x_{b2}$, and there exists a number   $\tilde{a}^* \in (x_{b1}, x_{b2})$ satisfying \eqref{eq:XOUa} such that
\begin{align}\label{XOU2cond1}
\frac{e^{\tilde{a}^*}+c_b}{F(\tilde{a}^*)} < \frac{e^{b^*} - c_s}{F(b^*)},
\end{align}
then the optimal switching problems \eqref{J} and \eqref{V} admit the solutions
\begin{align}
\tilde{J}(x) &=\displaystyle\begin{cases}
\tilde{P}F(x) &\, \textrm{ if }\, x\in (-\infty, \tilde{a}^*),\\
\tilde{K}F(x) -(e^x+c_b) &\, \textrm{ if }\, x\in [\tilde{a}^*, \tilde{d}^*], \\
\tilde{Q}G(x) &\, \textrm{ if }\, x\in (\tilde{d}^*, +\infty),
\end{cases}\label{XOUJsol}\\
\tilde{V}(x) &=\displaystyle \begin{cases}
\tilde{K}F(x) &\, \textrm{ if }\, x\in (-\infty, \tilde{b}^*),\\
\tilde{Q}G(x) + e^x-c_s &\, \textrm{ if }\, x\in [\tilde{b}^*, +\infty),
\end{cases} \label{XOUVsol}
\end{align}
where $\tilde{a}^*$ satisfies \eqref{eq:XOUa}, and
\begin{align}
\tilde{P}= \tilde{K}-\frac{e^{\tilde{a}^*}+c_b}{F(\tilde{a}^*)},  \quad
\tilde{K}=\frac{e^{\tilde{d}^*}G(\tilde{d}^*)-(e^{\tilde{d}^*}+c_b)G'(\tilde{d}^*)}{F'(\tilde{d}^*)G(\tilde{d}^*)-F(\tilde{d}^*)G'(\tilde{d}^*)},  \quad \tilde{Q}=\frac{e^{\tilde{d}^*}F(\tilde{d}^*)-(e^{\tilde{d}^*}+c_b)F'(\tilde{d}^*)}{F'(\tilde{d}^*)G(\tilde{d}^*)-F(\tilde{d}^*)G'(\tilde{d}^*)}.\label{QXOU}
\end{align}
There exist unique critical levels $\tilde{d}^*$ and  $\tilde{b}^*$ which are found from the nonlinear system of equations:
\begin{align}
\frac{e^{d}G(d)-(e^{d}+c_b)G'(d)}{F'(d)G(d)-F(d)G'(d)}=\frac{e^{b}G(b)-(e^{b}-c_s)G'(b)}{F'(b)G(b)-F(b)G'(b)},\label{eq:solveXOUG}\\
\frac{e^{d}F(d)-(e^{d}+c_b)F'(d)}{F'(d)G(d)-F(d)G'(d)}=\frac{e^{b}F(b)-(e^{b}-c_s)F'(b)}{F'(b)G(b)-F(b)G'(b)}.\label{eq:solveXOUF}
\end{align}
Moreover, the critical levels are such that $\tilde{d}^* \in (x_{b1}, x_{b2})$ and $\tilde{b}^* > x_s$. 
\end{theorem}

The optimal strategy  in Theorem \ref{thm:XOU2} is described by the stopping times $\Lambda_0^*=(\nu_1^*,\tau_1^*,\nu_2^*,\tau_2^*,\dots)$, and $\Lambda_1^*=(\tau_1^*,\nu_2^*,\tau_2^*,\nu_3^*,\dots)$, with  
\begin{align}\label{swtime}
&\nu_1^* = \inf\{t\geq 0: X_t\in [\tilde{a}^*, \tilde{d}^*]\},\\
&\tau_i^* = \inf\{t\geq \nu_i^*: X_t \geq \tilde{b}^*\}, \quad \textrm{and} \quad \nu_{i+1}^*=\inf\{t\geq \tau_i^*: X_t \leq \tilde{d}^*\}, \quad \textrm{for } i \geq 1.
\end{align}
In other words, it is optimal to buy if the price is within $[e^{\tilde{a}^{*}},e^{\tilde{d}^{*}}]$ and then  sell when the price $\xi$ reaches  $e^{\tilde{b}^*}$. The structure of the buy/sell regions is similar to that in the double stopping case (see Theorems \ref{thm:optLiquexpOU} and \ref{thm:optEntryexpOU}). In particular, $\tilde{a}^*$ is the same as $a^*$ in Theorem \ref{thm:optEntryexpOU} since the equations \eqref{eq:solveaexpOU} and \eqref{eq:XOUa} are equivalent. The level $\tilde{a}^*$ is only relevant to the first purchase. Mathematically,  $\tilde{a}^*$ is determined separately from $\tilde{d}^*$ and $\tilde{b}^*$.  If we start with a zero position, then it is optimal to enter if the price $\xi$ lies  in the interval $[e^{\tilde{a}^{*}},e^{\tilde{d}^{*}}]$. However, on all subsequent trades, we enter as soon as the price hits $e^{\tilde{d}^{*}}$ from above (after exiting at  $e^{\tilde{b}^{*}}$ previously). Hence, the lower level $\tilde{a}^*$ becomes irrelevant after the first entry.

Note that the conditions that differentiate  Theorems \ref{thm:XOU1} and \ref{thm:XOU2} are exhaustive and mutually exclusive. If the conditions in Theorem \ref{thm:XOU1} are violated, then the conditions in Theorem \ref{thm:XOU2} must hold. In particular, condition \eqref{eq:XOUa} in Theorem \ref{thm:XOU1} holds if and only if 
\begin{align}\label{XOUcond2}
\left\lvert\int_{-\infty}^{x_{b1}} \Psi(x)e^xf_b(x)dx \right\rvert <\int_{x_{b1}}^{x_{b2}} \Psi(x)e^xf_b(x)dx,
\end{align}
where
\begin{align}\label{Psi}
\Psi(x) = \frac{2F(x)}{\sigma^2\W(x)}, \quad \textrm{and} \quad \W(x) = F'(x)G(x)-F(x)G'(x) >0.
\end{align}  
Inequality  \eqref{XOUcond2} can be numerically verified given the model inputs.  


\subsection{Numerical Examples}
\begin{figure}[ht]
\begin{center}\includegraphics[width=3.1in]{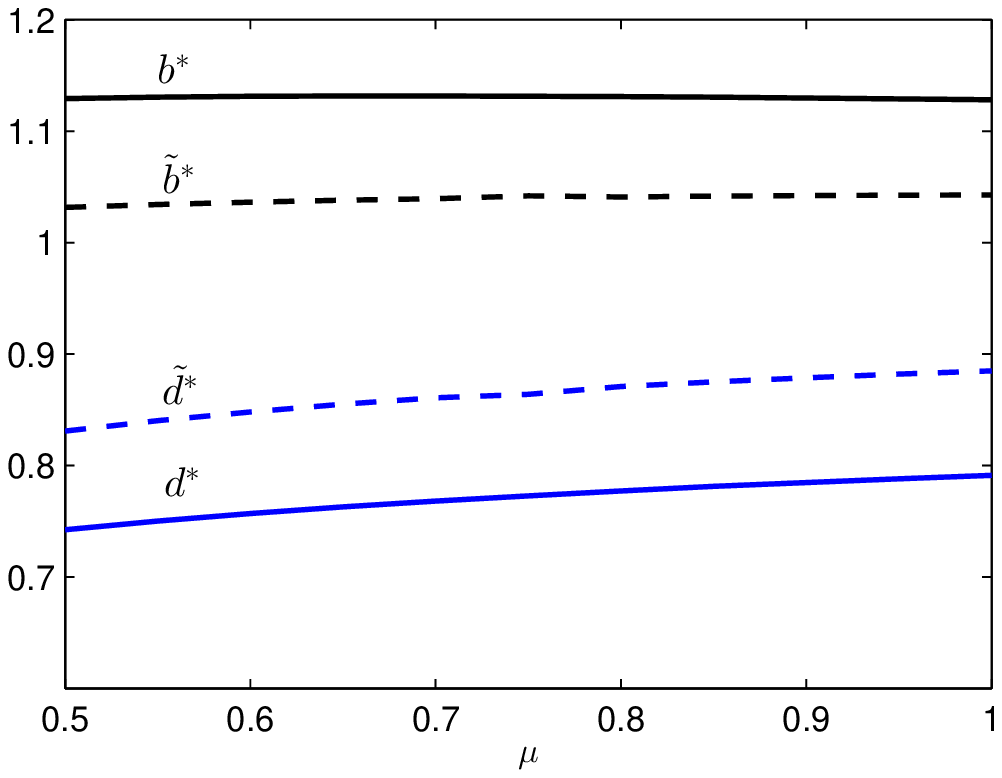}
\includegraphics[width=3.1in]{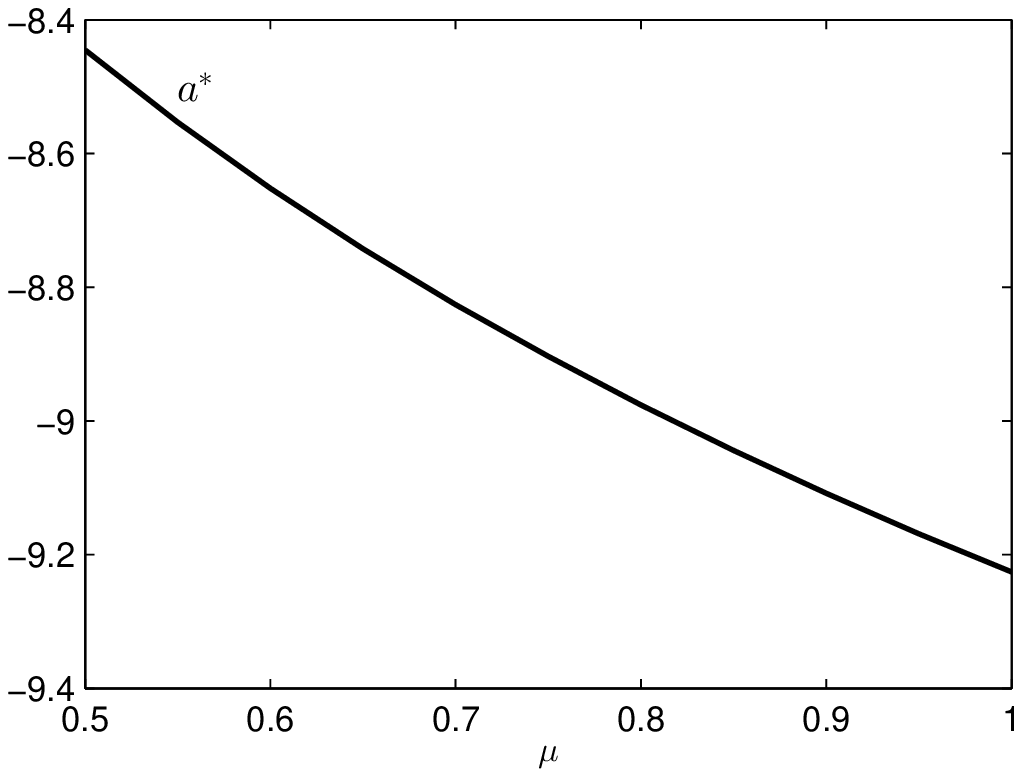}

\caption{\small{(Left) The optimal entry and exit levels vs speed of mean reversion $\mu$.  Parameters: $\sigma = 0.2$, $\theta = 1$, $r=0.05$, $c_s=0.02$, $c_b=0.02$.  (Right) The critical lower level of entry region $a^*$ decreases monotonically from -8.4452 to -9.2258 as $\mu$ increases from $0.5$ to $1.$ Parameters: $\sigma = 0.2$, $\theta = 1$, $r=0.05$, $c_s=0.02$, $c_b=0.02$.   }}
\label{fig:EOU_mu}
\end{center}\end{figure}

We numerically implement Theorems \ref{thm:optLiquexpOU}, \ref{thm:optEntryexpOU}, and \ref{thm:XOU2}, and illustrate the associated entry/exit thresholds.  In Figure \ref{fig:EOU_mu} (left), the optimal entry levels $d^*$ and $\tilde{d}^*$ rise, respectively, from 0.7425  to 0.7912 and  from 0.8310 to 0.8850, as the speed of mean reversion  $\mu$ increases from 0.5 to 1.  On the other hand, the critical exit levels  $b^*$ and  $\tilde{b}^*$ remain relatively flat over  $\mu$. As for the critical lower level $a^*$ from the  optimal double stopping problem,   Figure \ref{fig:EOU_mu} (right) shows that it is decreasing in $\mu$. The same pattern holds for the  optimal switching problem since  the critical lower level  $\tilde{a}^*$ is identical  to   $a^*$, as noted above.

\begin{figure}[ht]
\begin{center}\includegraphics[width=3.1in]{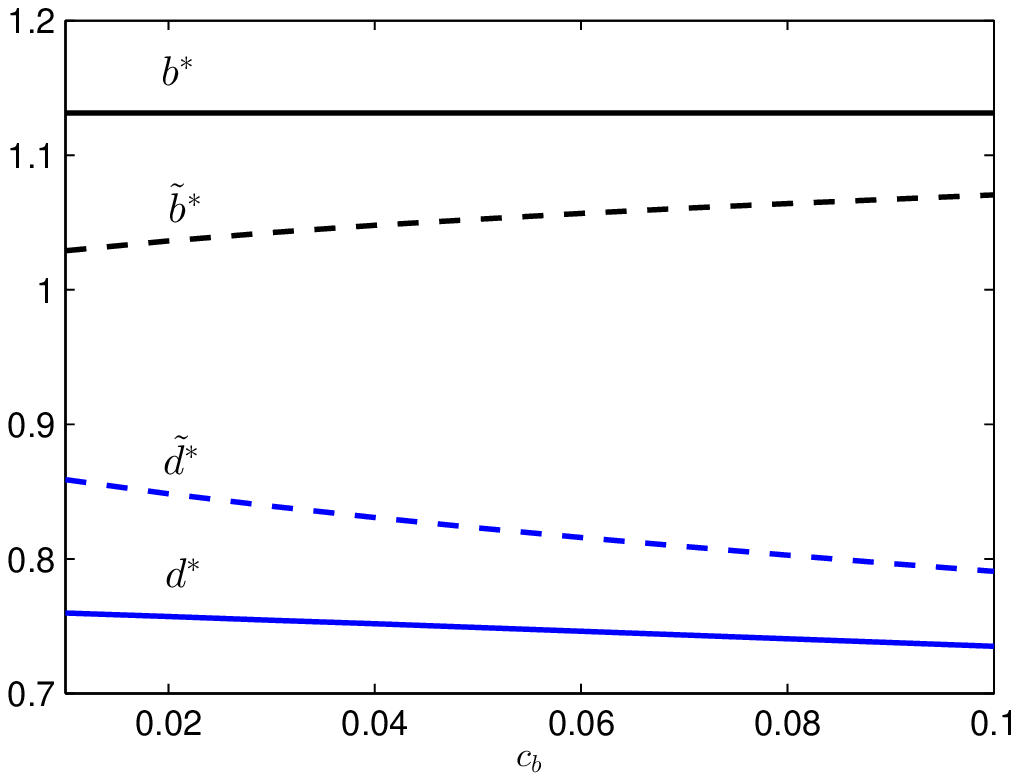}
\includegraphics[width=3.1in]{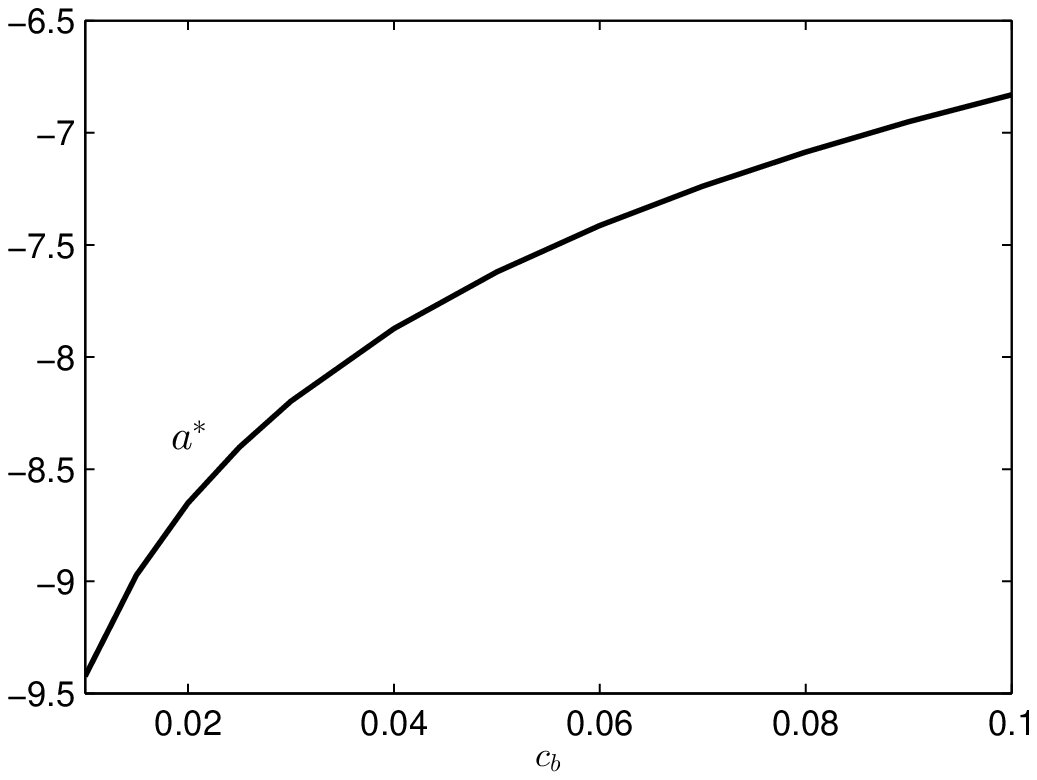}
\caption{\small{(Left) The optimal entry and exit levels vs transaction cost $c_b$. Parameters: $\mu = 0.6$, $\sigma = 0.2$, $\theta = 1$, $r=0.05$, $c_s=0.02$.  (Right) The  critical lower level of entry region $a^*$  increases monotonically from -9.4228 to -6.8305 as  $c_b$ increases from $0.01$ to $0.1.$ Parameters: $\mu = 0.6$, $\sigma = 0.2$, $\theta = 1$, $r=0.05$, $c_s=0.02$.   }}
\label{fig:EOU_cb}
\end{center}\end{figure}

We now look at the impact of transaction cost  in Figure \ref{fig:EOU_cb}. On the left panel,  we observe that as the transaction cost $c_b$ increases, the gap between the optimal switching entry and exit levels,  $\tilde{d}^*$ and $\tilde{b}^*$,  widens. This means that  it is optimal to  delay both entry and exit.  Intuitively, to counter the fall in profit margin due to an increase in transaction cost, it is necessary to buy at a lower price and sell at a higher price to seek a wider spread. In comparison, the exit level $b^*$ from  the double stopping problem is known   analytically  to be independent of the entry cost, so it stays constant as $c_b$ increases in the figure. In contrast, the entry level $d^*$, however, decreases  as  $c_b$ increases but much less significantly than  $\tilde{d}^*$.  Figure \ref{fig:EOU_cb} (right) shows that $a^*$, which is the same for   both the optimal double stopping and switching problems, increases monotonically  with $c_b$. \\

In both  Figures \ref{fig:EOU_mu} and \ref{fig:EOU_cb}, we can see that the interval of  the entry and exit levels,  $(\tilde{d}^*, \tilde{b}^*)$, associated with the optimal switching problem lies within the corresponding interval $(d^*, b^*)$ from the optimal  double stopping problem. Intuitively,  with the intention to enter the market again upon completing the current trade, the trader is more willing to enter/exit earlier, meaning a  narrowed waiting  region.  \\

\begin{figure}[t]
\begin{center}
\includegraphics[width=4in]{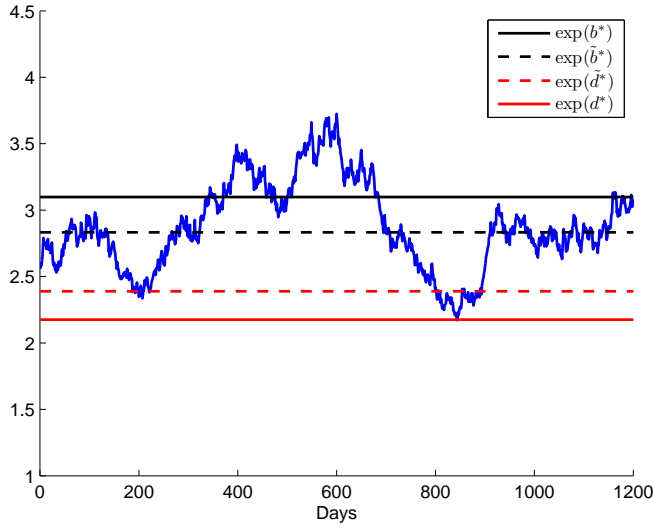}
\caption{\small{A sample exponential OU path, along with entry and exit levels. Under the double stopping setting, the investor enters at $\nu_{d^*} = \inf\{t\geq 0: \xi_t \leq e^{d^*} = 2.1754\}$ with $d^* = 0.7772$, and exit at $\tau_{b^*} = \inf\{ t\geq \nu_{d^*}: \xi_t \geq e^{b^*}= 3.0988\}$ with $b^* = 1.1310$.  The optimal switching investor enters at $\nu_{\tilde{d}^*} = \inf\{t\geq 0: \xi_t \leq e^{\tilde{d}^*} = 2.3888\}$ with $\tilde{d}^* = 0.8708$, and exit at $\tau_{\tilde{b}^*} = \inf\{ t\geq \nu_{\tilde{d}^*}: \xi_t \geq e^{\tilde{b}^*} = 2.8323\}$ with $\tilde{b}^* = 1.0411$. The critical lower threshold of entry region is $e^{a^*} = 1.264\cdot 10^{-4}$ with  $a^* = -8.9760$ (not shown in this figure). Parameters: $\mu = 0.8$, $\sigma = 0.2$, $\theta = 1$, $r=0.05$, $c_s=0.02$, $c_b=0.02$.   }}
\label{fig:EOU_Sim_1}
\end{center}\end{figure}

Figure \ref{fig:EOU_Sim_1} shows a simulated path and the associated entry/exit levels. As the path starts at  $\xi_0 = 2.6011 > e^{\tilde{d}^*} > e^{d^*}$, the investor waits to enter until the path reaches the lower level $e^{d^*}$  (double stopping) or $e^{\tilde{d}^*}$ (switching) according to  Theorems \ref{thm:optEntryexpOU} and  \ref{thm:XOU2}. After entry, the investor exits at the optimal level $e^{b^*}$  (double stopping) or $e^{\tilde{b}^*}$ (switching).    The  optimal switching thresholds   imply that  the investor first enters the market on day 188 where the underlying asset price is $2.3847$. In contrast,  the optimal double stopping timing yields  a later entry   on day 845  when the price first reaches $e^{d^*} = 2.1754$.  As for the  exit timing, under the optimal switching setting, the investor exits the market earlier on day 268 at  the price  $e^{\tilde{b}^*} = 2.8323$.  The double stopping timing  is  much later on day 1160 when the price reaches $e^{b^*} = 3.0988$.  In addition, under the optimal switching problem, the investor   executes more  trades within the same time span. As seen in  the figure, the investor would have completed two `round-trip' (buy-and-sell) trades in the market    before the double stopping investor liquidates for the first time.

\section{Methods of Solution and Proofs}\label{sect-method}
We now provide detailed proofs for our analytical results in Section \ref{sect-solution} beginning with Theorems \ref{thm:optLiquexpOU} and \ref{thm:optEntryexpOU} for the optimal double stopping problems.

\subsection{Optimal Double Stopping Problems}\label{sect-XOUexitproof}

Starting at  any $x\in \R$, we denote  by  $\tau_a\wedge\tau_b$  the exit time from an  interval $[a,b]$ with  $-\infty \le a \le x \le b \le +\infty$. If  $a=-\infty$, then we have $\tau_a = +\infty$ a.s. In effect, this removes the  lower exit level. Similarly, it is possible that $b=+\infty$, and  $\tau_b = +\infty$ a.s. Consequently, by considering interval-type strategies, we also include the class of stopping strategies  of reaching a single  level (as in Theorem \ref{thm:optLiquexpOU} above).

Now, let us introduce   the  transformation  \begin{align}\label{psi2}\psi(x):=  \frac{F}{G}(x),\end{align} and define   $z_a=\psi(a)$, $z_b=\psi(b)$.  With  the reward function $h_s(x)$,  and using \eqref{psi2}, we compute the corresponding expected discounted reward:
\begin{align}
\E_x\{e^{-r(\tau_a\wedge\tau_b)}h_s(X_{\tau_a\wedge\tau_b})\} &= h_s(a)\E_x\{e^{-r(\tau_a\wedge\tau_b)}\indic{\tau_a<\tau_b}\} + h_s(b) \E_x\{e^{-r(\tau_a\wedge\tau_b)}\indic{\tau_a>\tau_b}\}\label{EH1}\\
&= h_s(a) \frac{F(x)G(b)-F(b)G(x)}{F(a)G(b)-F(b)G(a)} +  h_s(b)\frac{F(a)G(x)-F(x)G(a)}{F(a)G(b)-F(b)G(a)} \label{EH12}\\
&= G(x)\left[\frac{h_s(a)}{G(a)}\frac{\psi(b)-\psi(x)}{\psi(b)-\psi(a)} + \frac{h_s(b)}{G(b)}\frac{\psi(x)-\psi(a)}{\psi(b)-\psi(a)} \right]\label{EH2}\\
&= G(\psi^{-1}(z))\left[H(z_a)\frac{z_b-z}{z_b-z_a}+H(z_b)\frac{z-z_a}{z_b-z_a} \right], \label{EH3}
\end{align}
where 
\begin{align}\label{generalH}
H(z)  := \begin{cases}
\frac{h_s}{G}\circ \psi^{-1}(z) &\, \textrm{ if }\, z>0,\\
\lim_{x\to -\infty}\limits\frac{(h_s(x))^+}{G(x)} &\, \textrm{ if }\, z=0.
\end{cases}
\end{align}
The last equality \eqref{EH3} transforms the problem from $x$ coordinate to $z = \psi(x)$  coordinate (see \eqref{psi2}).

In turn, the candidate optimal exit interval $[a^*, b^*]$ is determined by maximizing the expectation in \eqref{EH1}. This is equivalent to maximizing \eqref{EH3} over $z_{a}$ and $z_{b}$  in  the transformed problem. As a result, for every $z \geq 0,$ we have
\begin{align}
W(z) := \sup_{\{z_a,z_b: z_a\leq z\leq z_b\}} \left\{H(z_a)\frac{z_b-z}{z_b-z_a}+H(z_b)\frac{z-z_a}{z_b-z_a}\right\},\label{Wy1}
\end{align}
which is the smallest concave majorant of $H$. Applying \eqref{Wy1}  to  \eqref{EH3}, we can express the maximal expected discounted reward as
\[G(x)W(\psi(x)) =\sup_{\{a,b: a \leq x \leq b\}} \E_x\{e^{-r(\tau_a\wedge\tau_b)}h_s(X_{\tau_a\wedge\tau_b})\}. \]

Now,  it remains to prove the optimality of the proposed stopping strategy. This also  provides an analytic  expression for the value function.

\begin{theorem}\label{thm:V}
Under the XOU model \eqref{XOU}, the value function $V(x)$ defined in \eqref{V1a} is given by
\begin{align}\label{generalV}
V(x) = G(x)W(\psi(x)),
\end{align}
where $G$, $\psi$ and $W$ are defined in \eqref{GOU}, \eqref{psi2} and \eqref{Wy1}, respectively.
\end{theorem}

The proof is similar to that of Theorem 3.2 in \cite{LeungLi2014OU}, and is thus omitted.

 By  Theorem \ref{thm:V}, it is sufficient to consider interval-type strategies for the optimal liquidation problem under the XOU model. Note that the optimal levels $(a^*,  b^*)$ can  depend on the initial value  $x$,  and they may coincide or  take values $-\infty$ or $+\infty$. As such,  the structure of the stopping and continuation regions can potentially  be characterized by multiple intervals, leading to disconnected  continuation regions (see Theorem \ref{thm:optEntryexpOU} above).   In order to determine the optimal exit timing strategies and solve for $V$, the major challenge  lies in analyzing the  functions $H$ and  $W$.

\subsubsection{Optimal Exit Timing}\label{sect-XOUexitproof}
In preparation for the next result,  we apply  \eqref{psi2} and \eqref{FOU}-\eqref{GOU}  to the definition  of   $H$ in  \eqref{generalH}, and  summarize the crucial properties of $H$.

\begin{lemma}\label{lm:HXOU}
The function $H$ is continuous on $[0,+\infty)$, twice differentiable on $(0,+\infty)$ and possesses the following properties:
\begin{enumerate}[(i)]
\item \label{HXOU0} $H(0)=0$, and
\begin{align*}
H(z) \begin{cases}
<0 &\, \textrm{ if }\, z\in (0,\psi(\ln c_s)),\\
>0 &\, \textrm{ if }\, z\in (\psi(\ln c_s),+\infty).
\end{cases}
\end{align*}

\item \label{HXOU1}
$H(z)$ is strictly increasing for $z \in (\psi(\ln c_s), +\infty)$, and $H'(z)\to 0$ as $z\to +\infty$.

\item \label{HXOU2}
\begin{align*}
H(z) \textrm{ is }\begin{cases}
\textrm{convex} &\, \textrm{ if }\, z\in (0,\psi(x_s)],\\
\textrm{concave} &\, \textrm{ if }\, z\in [\psi(x_s),+\infty).
\end{cases}
\end{align*}
\end{enumerate}
\end{lemma}

\begin{figure}[th!]
\begin{center}
 {\scalebox{0.35}{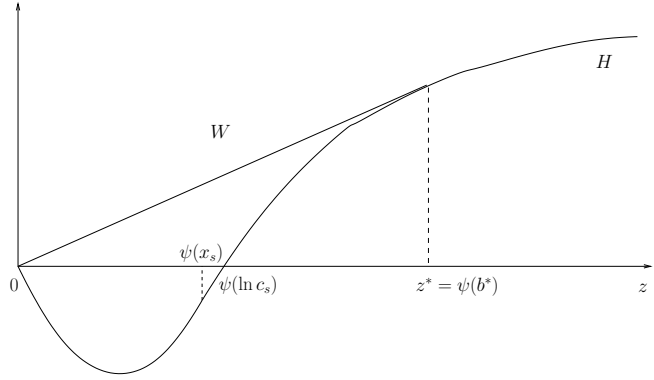}}
\end{center}
\caption{{\small Sketches  of $H$ and $W$. By Lemma \ref{lm:HXOU}, $H$ is convex on the left of  $\psi(x_s)$ and concave on the right. The smallest concave majorant  $W$ is a straight line  tangent to $H$ at $z^*$ on $[0,z^*)$, and coincides with $H$ on $[z^*,+\infty)$. }}
\label{fig:HXOU}
\end{figure}

Based on Lemma \ref{lm:HXOU}, We sketch $H$ in Figure \ref{fig:HXOU}. Using the properties of $H$, we now solve for the optimal exit timing.

\paragraph*{Proof of Theorem \ref{thm:optLiquexpOU}}
We look for the value function of the form: $V(x)  = G(x)W(\psi(x))$, where $W$ is the smallest concave majorant of $H$. By  Lemma \ref{lm:HXOU}, we observe that  $H$ is concave over  $[\psi(x_s),+\infty)$, strictly positive over  $(\psi(\ln c_s),+\infty)$,  and ${H}^\prime\!(z) \to 0$ as $z\to +\infty$.
Therefore,  there exists a unique number
$z^*>\psi(x_s) \vee \psi(\ln c_s)$ such that
\begin{align}\label{eq:HzLiquXOU}
\frac{H(z^*)}{z^*}={H}^\prime\!(z^*).
\end{align}
In turn, the smallest concave majorant of $H$ is given by
\begin{align*}
W(z) = \begin{cases}
z\frac{H(z^*)}{z^*} &\, \textrm{ if }\, z \in [0, z^*),\\
H(z) &\, \textrm{ if }\, z \in [z^*,+\infty).
\end{cases}
\end{align*}
Substituting $b^{*} = \psi^{-1}(z^*)$ into \eqref{eq:HzLiquXOU},
we have
\[ \frac{H(z^*)}{z^*} = \frac{H(\psi(b^{*}))}{\psi(b^{*})} = \frac{e^{b^{*}}-c_s}{F(b^{*})},\]
and
\begin{align*}
{H}^\prime\!(z^*) &= \frac{e^{\psi^{-1}(z^*)}G(\psi^{-1}(z^*))-(e^{\psi^{-1}(z^*)}-c_s)G'(\psi^{-1}(z^*))}{F'(\psi^{-1}(z^*))G(\psi^{-1}(z^*)) - F(\psi^{-1}(z^*)) G'(\psi^{-1}(z^*))}\\
&= \frac{e^{b^{*}}G(b^{*}) - (e^{b^{*}}-c_s)G'(b^{*})}{F'(b^{*})G(b^{*})-F(b^{*})G'(b^{*})}.
\end{align*}
Equivalently, we can express \eqref{eq:HzLiquXOU} in terms of
$b^{*}$:
\begin{align*}
\frac{e^{b^{*}}-c_s}{F(b^{*})} = \frac{e^{b^{*}}G(b^{*}) - (e^{b^{*}}-c_s)G'(b^{*})}{F'(b^{*})G(b^{*})-F(b^{*})G'(b^{*})},
\end{align*}
which is equivalent to \eqref{eq:solvebexpOU} after simplification.
As a result, we have
\begin{align*}
W(\psi(x)) = \begin{cases}
\psi(x) \frac{H(z^*)}{z^*}  = \frac{F(x)}{G(x)}\frac{e^{b^{*}}-c_s}{F(b^{*})} &\, \textrm{ if }\, x \in (-\infty, b^{*}),\\
H(\psi(x)) = \frac{e^{x}-c_s}{G(x)} &\, \textrm{ if }\, x \in [b^{*},+\infty).
\end{cases}
\end{align*}
In turn, the value function $V(x)  = G(x)W(\psi(x))$ is given by \eqref{VexpOUsol}.

\subsubsection{Optimal Entry Timing}\label{sect-XOUentryproof}
We can directly  follow the arguments that yield Theorem \ref{thm:V}, but with  the reward as $\hat{h}(x)=V(x) - h_b(x) = V(x) - (e^x+c_b)$ and define $\hat{H}$  analogous to  $H$:
\begin{align}\label{generalhatH}
\hat{H}(z)  := \begin{cases}
\frac{\hat{h}}{G}\circ \psi^{-1}(z) &\, \textrm{ if }\, z>0,\\
\lim_{x\to -\infty}\limits\frac{(\hat{h}(x))^+}{G(x)} &\, \textrm{ if }\, z=0.
\end{cases}
\end{align}
We will look for the value function with the form: $J(x) = G(x)\hat{W}(\psi(x))$, where $\hat{W}$ is the smallest concave majorant of $\hat{H}$. The properties of $\hat{H}$ is given in the next lemma.

\begin{lemma}\label{lm:hatHXOUr}
The function $\hat{H}$ is continuous on $[0,+\infty)$, differentiable
on $(0,+\infty)$, and twice differentiable on $(0,\psi(b^{*})) \cup
(\psi(b^{*}),+\infty)$, and possesses the following properties:
\begin{enumerate}[(i)]
\item \label{hatHXOU0r} $\hat{H}(0)=0$, and there exists some $\underline{b} <
    b^{*}$ such that $\hat{H} (z)<0$ for $z \in (0,
    \psi(\underline{b}))\cup [\psi(b^{*}), +\infty)$.

\item  \label{hatHXOU1r} $\hat{H} (z)$ is strictly decreasing for $z \in
    [\psi(b^{*}), +\infty)$.

\item \label{hatHXOU2r} Define the constant \[x^{*} = \theta+\frac{\sigma^2}{2\mu}-\frac{r}{\mu}-1.\]
There exist some constants $x_{b1}$ and $x_{b2}$,  with $-\infty < x_{b1} < x^{*} < x_{b2} < x_s$, that solve $f_b(x)=0$, such that
\begin{align*}
\hat{H} (z) \textrm{ is }
\begin{cases}
\textrm{convex} &\, \textrm{ if }\, z \in (0, \psi(x_{b1}))\cup (\psi(x_{b2}),+\infty)\\
\textrm{concave} &\, \textrm{ if }\, z \in (\psi(x_{b1}),\psi(x_{b2})),
\end{cases}
\end{align*}
and  $\hat{z}_1 := \argmax_{z\in [0,+\infty)} \hat{H}(z) \in (\psi(x_{b1}),\psi(x_{b2}))$.
\end{enumerate}
\end{lemma}

\begin{figure}[h]
\begin{center}
 {\scalebox{0.35}{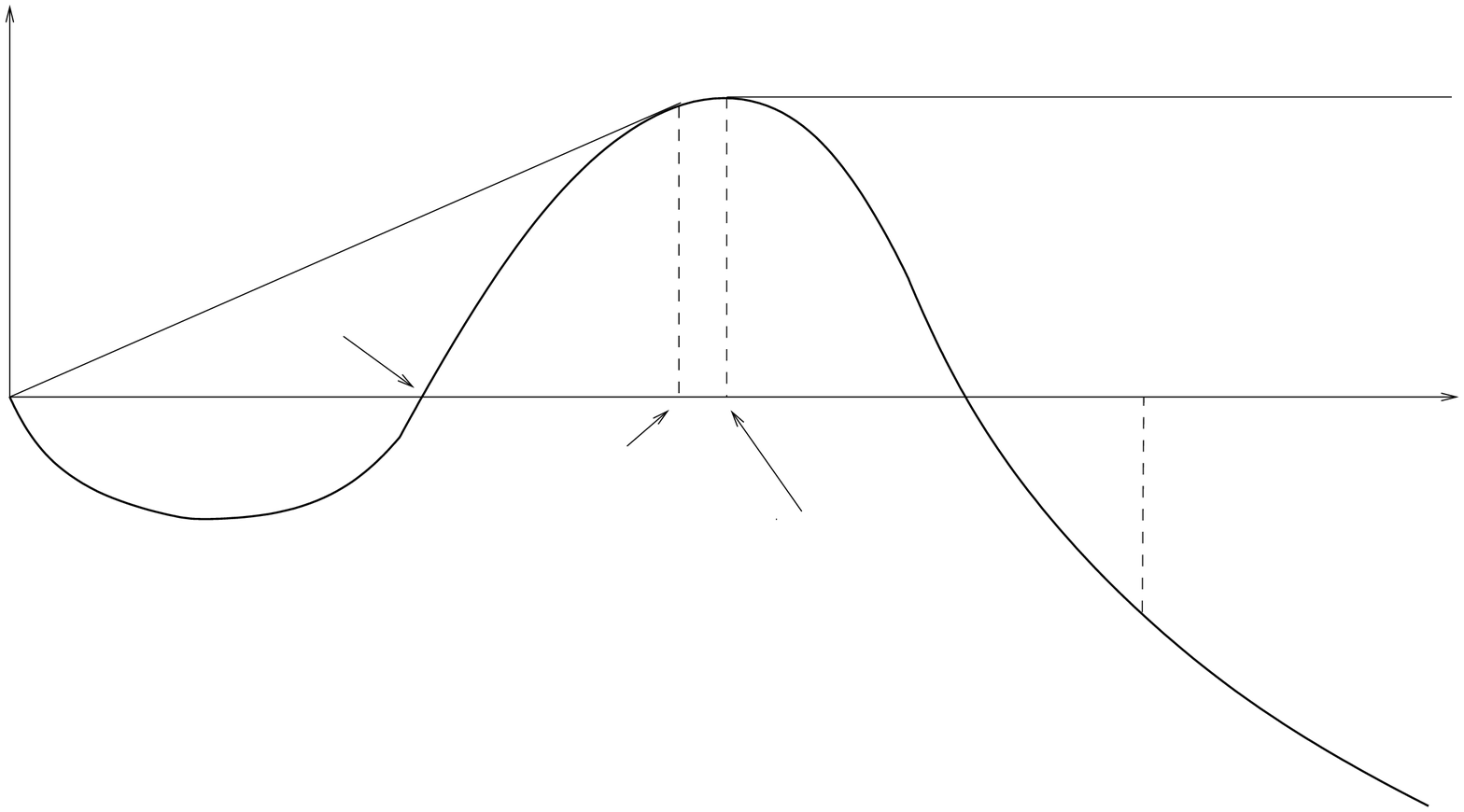}}
\end{center}
\setlength{\abovecaptionskip}{-5pt}
\caption{\small{Sketches of $\hat{H}$ and $\hat{W}$. The smallest concave majorant $\hat{W}$ is a straight line tangent to $\hat{H}$ at $\hat{z}_0$ on $[0,\hat{z}_0)$, coincides with $\hat{H}$ on $[\hat{z}_0,\hat{z}_1]$, and is equal to $\hat{H}(\hat{z}_1)$ on $(\hat{z}_1,+\infty)$. }}
\label{fig:hatHXOU}
\end{figure}

 Figure \ref{fig:hatHXOU} gives a sketch of  $\hat{H}$  according to Lemma \ref{lm:hatHXOUr}, and illustrate the corresponding smallest concave majorant  $\hat{W}$.

\paragraph*{Proof of Theorem \ref{thm:optEntryexpOU}}
 As in Lemma \ref{lm:hatHXOUr} and Figure \ref{fig:hatHXOU}, by the definition of the maximizer of $\hat{H}$, $\hat{z}_1$ satisfies the equation
\begin{align}\label{eq:Hz1EntryXOU}
\hat{H}^{'}\!(\hat{z}_1)=0.
\end{align}
Also there exists a unique number $\hat{z}_0 \in (x_{b1},\hat{z}_1)$ such that
\begin{align}\label{eq:Hz0EntryXOU}
\frac{\hat{H}(\hat{z}_0)}{\hat{z}_0} =\hat{H}^{'}\!(\hat{z}_0).
\end{align}

Using \eqref{eq:Hz1EntryXOU}, \eqref{eq:Hz0EntryXOU} and Figure \ref{fig:hatHXOU},  $\hat{W}$ is  a straight line tangent to $\hat{H}$ at $\hat{z}_0$ on $[0,\hat{z}_0)$, coincides with $\hat{H}$ on $[\hat{z}_0,\hat{z}_1]$, and is equal to $\hat{H}(\hat{z}_1)$ on $(\hat{z}_1,+\infty)$. As a result,
\begin{align*}
\hat{W}(z) = \begin{cases}
z\hat{H}^{'}\!(\hat{z}_0) &\, \textrm{ if }\, z\in [0,\hat{z}_0),\\
\hat{H}(y) &\, \textrm{ if }\, z\in [\hat{z}_0, \hat{z}_1],\\
\hat{H}(\hat{z}_1) &\, \textrm{ if }\, z\in (\hat{z}_1, +\infty).
\end{cases}
\end{align*}
Substituting $a^{*} = \psi^{-1}(\hat{z}_0)$ into
\eqref{eq:Hz0EntryXOU}, we have
\begin{align*}
\frac{\hat{H}(\hat{z}_0)}{\hat{z}_0} =\frac{V(a^{*})-(e^{a^{*}}+c_b)}{F(a^{*})},
\end{align*}
and
\begin{align*}
\hat{H}^{'}\!(\hat{z}_0)= \frac{G(a^{*})(V'(a^{*})  - e^{a^{*}}) - G'(a^{*})(V(a^{*})-(e^{a^{*}}+c_b))}
{F'(a^{*})G(a^{*})-F(a^{*})G'(a^{*})}.
\end{align*}
Equivalently, we can express condition \eqref{eq:Hz0EntryXOU} in terms
of $a^{*}$:
\begin{align*}
\frac{V(a^{*})-(e^{a^{*}}+c_b)}{F(a^{*})} = \frac{G(a^{*})(V'(a^{*})  - e^{a^{*}}) - G'(a^{*})(V(a^{*})-(e^{a^{*}}+c_b))}
{F'(a^{*})G(a^{*})-F(a^{*})G'(a^{*})},
\end{align*}
which  is equivalent to \eqref{eq:solveaexpOU} after simplification.  Also, we can express
$\hat{H}^{'}\!(\hat{z}_0)$ in terms of $a^{*}$:
\begin{align*}
\hat{H}^{'}\!(\hat{z}_0)=\frac{\hat{H}(\hat{z}_0)}{\hat{z}_0} = \frac{V(a^{*})-(e^{a^{*}}+c_b)}{F(a^{*})} = P.
\end{align*}
In addition, substituting $d^{*} = \psi^{-1}(\hat{z}_1)$
into \eqref{eq:Hz1EntryXOU}, we have
\begin{align*}
\frac{G(d^{*})(V'(d^{*})  - e^{d^{*}}) - G'(d^{*})(V(d^{*})-(e^{d^{*}}+c_b))}
{F'(d^{*})G(d^{*})-F(d^{*})G'(d^{*})}=0,
\end{align*} which can be further simplified to  \eqref{eq:solvedexpOU}.
Furthermore,
$\hat{H}(\hat{z}_1)$ can be written in terms of $d^{*}$:
\begin{align*}
\hat{H}(\hat{z}_1) = \frac{V(d^{*})-(e^{d^{*}}+c_b)}{G(d^{*})}=Q.
\end{align*}
By direct substitution of  the expressions for $\hat W$ and the associated functions, we obtain the value function in \eqref{JsoluexpOU}.

 \subsection{Optimal Switching Problems}\label{sect-XOUswproof}
Using the results derived in previous sections, we can infer  the structure of the buy and sell regions of the switching problem and then proceed to verify  its optimality. In this section, we provide detailed proofs for  Theorems \ref{thm:XOU1} and \ref{thm:XOU2}.
\paragraph*{Proof of Theorem \ref{thm:XOU1} (Part 1)}\label{pf:XOU1}
First, with $h_s(x)=e^x-c_s$, we differentiate to get
\begin{align}\label{hsF1}
\left(\frac{h_s}{F}\right)'(x) = \frac{(e^x-c_s)F'(x)-e^x F(x)}{F^2(x)}.
\end{align}
On the other hand, by Ito's lemma, we have
\begin{align*}
h_s(x) = \E_x\{e^{-rt}h_s(X_t)\} - \E_x\left\{\int_0^t e^{-ru}(\L-r)h_s(X_u)du\right\}.
\end{align*}
Note that
\begin{align*}
\E_x\{e^{-rt}h_s(X_t)\}= e^{-rt}\left(e^{(x-\theta)e^{-\mu t}+\theta+\frac{\sigma^2}{4\mu}(1-e^{-2\mu t})}-c_s \right) \to 0 \quad \textrm{as} \quad t\to +\infty.
\end{align*}
This implies that 
\begin{align}
h_s(x) &= -\E_x\left\{\int_0^{+\infty} e^{-ru}(\L-r)h_s(X_u)du\right\}\notag\\
&=-G(x)\int_{-\infty}^x \Psi(s) (\L-r) h_s(s)ds - F(x)\int_x^{+\infty} \Phi(s)(\L-r) h_s(s)ds,\label{hsrep}
\end{align}
where $\Psi$ is defined in \eqref{Psi} and
\begin{align*}
\Phi(x):= \frac{2G(x)}{\sigma^2\W(x)}.
\end{align*}
The last line follows from Theorem 50.7 in \citet[p.~293]{Rogers2000}.
Dividing both sides by $F(x)$ and differentiating the RHS of \eqref{hsrep}, we obtain
\begin{align*}
\left(\frac{h_s}{F}\right)'(x)&= -\left(\frac{G}{F} \right)'(x) \int_{-\infty}^x \Psi(s) (\L-r) h_s(s)ds  - \frac{G}{F}(x)\Psi(x) (\L-r) h_s(x) - \Phi(x)(\L-r) h_s(x)\\
 &= \frac{\W(x)}{F^2(x)}\int_{-\infty}^x \Psi(s)(\L-r)h_s(s)ds = \frac{\W(x)}{F^2(x)}q(x),
\end{align*}
where 
\begin{align}
q(x) := \int_{-\infty}^x \Psi(s)(\L-r)h_s(s)ds.
\end{align}
Since $\W(x), F(x)>0$, we deduce that $\left(\frac{h_s}{F}\right)'(x)=0$ is equivalent to $q(x)=0$. Using \eqref{hsF1}, we now see that \eqref{eq:solvebexpOU} is equivalent to $q(b)=0$.
 
%
Next, it follows from  \eqref{Lr} that
\begin{align}\label{q1XOU}
q'(x) = \Psi(x)(\L-r)h_s(x) \begin{cases}
> 0 &\, \textrm{ if }\, x<  x_s,\\
< 0 &\, \textrm{ if }\, x >  x_s.
\end{cases}
\end{align}
This, together with the fact that $\lim_{x\to-\infty}q(x)=0$, implies that there exists a unique $b^*$ such that $q(b^*)=0$ if and only if $\lim_{x \to +\infty}q(x) < 0$. Next, we show that this inequality holds.  By the definition of $h_s$ and $F$, we have
\begin{align}
\frac{h_s(x)}{F(x)} = \frac{e^x - c_s}{F(x)} > 0 \quad \textrm{for }    x > \ln c_s, \qquad  \quad \lim_{x \to +\infty}\frac{h_s(x)}{F(x)} = 0, \notag\\
\left(\frac{h_s}{F}\right)'(x) = \frac{\W(x)}{F^2(x)}\int_{-\infty}^x \Psi(s)(\L-r)h_s(s)ds = \frac{\W(x)}{F^2(x)}q(x). \label{eq:XOUhsF1}
\end{align}

Since $q$ is strictly decreasing in $(x_s, +\infty)$, the above hold true if and only if $\lim_{x \to +\infty}q(x) < 0$. Therefore, we conclude that there exits a unique $b^*$  such that $e^bF(b) = (e^b-c_s)F'(b)$. Using \eqref{q1XOU}, we see that
\begin{align}\label{XOUb0prop}
b^*>x_s \quad \textrm{and} \quad q(x) > 0 \quad \textrm{for all} \quad x < b^*.
\end{align}
Observing that $e^{b^*}, F(b^*), F'(b^*) >0$, we can conclude that $h_s(b^*)=e^{b^*}-c_s > 0$, or equivalently $b^* > \ln c_s$.

We now verify by direct substitution that $\tilde{V}(x)$ and $\tilde{J}(x)$ in \eqref{sol:XOU} satisfy the pair of variational inequalities:
\begin{align}
\min\{r\tilde{J}(x)-\L \tilde{J}(x), \tilde{J}(x) - (\tilde{V}(x) -h_b(x))\}&=0,\label{VIJsw}\\
\min\{r\tilde{V}(x)-\L \tilde{V}(x), \tilde{V}(x) - (\tilde{J}(x)+h_s(x))\}&=0.\label{VIVsw}
\end{align}
 First, note  that $\tilde{J}(x)$ is identically 0 and thus satisfies the equality
\begin{align}
(r-\L)\tilde{J}(x) = 0. \label{VIJsw1}
\end{align}
To show that $\tilde{J}(x)-(\tilde{V}(x)-h_b(x))\geq 0$, we look at the disjoint intervals $(-\infty, b^*)$ and $[b^*, \infty)$ separately. For $x \geq b^*,$ we have
\begin{align}
\tilde{V}(x) - h_b(x) = -(c_b + c_s),
\end{align}
which implies $\tilde{J}(x)-(\tilde{V}(x)-h_b(x)) = c_b + c_s \geq 0$. When $x < b^*,$ the inequality
\begin{align}
 \tilde{J}(x)-(\tilde{V}(x)-h_b(x))\geq 0
\end{align}
can be rewritten as
\begin{align}\label{XOU1sw}
\frac{h_b(x)}{F(x)} = \frac{e^x + c_b}{F(x)} \geq \frac{e^{b^*} - c_s}{F(b^*)} =\frac{h_s(b^*)}{F(b^*)}.
\end{align}
To determine the necessary conditions for this to hold, we consider the derivative of the LHS of \eqref{XOU1sw}:
\begin{align}\label{XOUhbF1}
\left(\frac{h_b}{F}\right)'(x) &= \frac{\W(x)}{F^2(x)}\int_{-\infty}^x \Psi(s)(\L-r)h_b(s)ds =  \frac{\W(x)}{F^2(x)}\int_{-\infty}^x \Psi(s) e^sf_b(s)ds.
\end{align}
If $f_b(x)=0$ has no roots, then $(\L-r)h_b(x)$ is negative for all $x \in \R$. On the other hand, if there is only one root $\tilde{x}$, then $(\L-r)h_b(\tilde{x}) = 0$ and $(\L-r)h_b(x) < 0$ for all other $x$. In either case, $h_b(x)/F(x)$ is a strictly decreasing function and \eqref{XOU1sw} is true.

Otherwise if $f_b(x)=0$ has two distinct roots $x_{b1}$ and $x_{b2}$ with $x_{b1} < x_{b2}$, then
\begin{align}\label{LrbXOU}
(\L-r)h_b(x)\begin{cases}
<0 &\, \textrm{ if }\, x\in (-\infty, x_{b1}) \cup (x_{b2}, +\infty),\\
>0 &\, \textrm{ if }\, x\in (x_{b1}, x_{b2}).
\end{cases}
\end{align}
Applying \eqref{LrbXOU} to \eqref{XOUhbF1}, the derivative $(h_b/F)'(x)$ is negative on $(-\infty, x_{b1})$. Hence, $h_b(x)/F(x)$ is strictly decreasing on $(-\infty, x_{b1})$. We further  note that $b^* \!>\! x_s \!>\! x_{b2}$. Observe that on the interval $(x_{b1}, x_{b2})$, the intergrand is positive. It is therefore possible for $(h_b/F)'$ to change sign at some $x \in (x_{b1}, x_{b2})$. For this to happen, the positive part of the integral must be larger than the absolute value of the negative part.  In other words, \eqref{XOUcond2} must hold. If \eqref{XOUcond2} holds, then there must exist some $\tilde{a}^* \in (x_{b1}, x_{b2})$ such that $(h_b/F)'(\tilde{a}^*) = 0$, or equivalently \eqref{eq:XOUa} holds:
\begin{align*}
\left(\frac{h_b}{F}\right)'(\tilde{a}^*) &= \frac{h_b'(\tilde{a}^*)}{F(\tilde{a}^*)} - \frac{h_b(\tilde{a}^*)F'(\tilde{a}^*)}{F^2(\tilde{a}^*)}= \frac{e^{\tilde{a}^*}}{F(\tilde{a}^*)} - \frac{(e^{\tilde{a}^*} + c_b)F(\tilde{a}^*)'}{F^2(\tilde{a}^*)}.
\end{align*}


If \eqref{eq:XOUa} holds, then we have 
\begin{align*}
\left\lvert\int_{-\infty}^{x_{b1}} \Psi(x) e^xf_b(x)dx\right\rvert = \int_{x_{b1}}^{\tilde{a}^*} \Psi(x) e^xf_b(x)dx. 
\end{align*}
In addition, since 
\begin{align*}
\int_{\tilde{a}^*}^{x_{b2}} \Psi(x) e^xf_b(x)dx > 0,
\end{align*}
it follows that
\begin{align*}
\left\lvert\int_{-\infty}^{x_{b1}} \Psi(x) e^xf_b(x)dx\right\rvert < \int_{x_{b1}}^{x_{b2}} \Psi(x) e^xf_b(x)dx.
\end{align*}
This establishes the equivalence between \eqref{eq:XOUa} and \eqref{XOUcond2}. Under this condition, $h_b/F$ is strictly decreasing on $(x_{b1},\tilde{a}^*)$. Then, either it is strictly increasing on $(\tilde{a}^*, b^*)$, or there exists some $\bar{x} \in  (x_{b2}, b^*)$ such that $h_b(x)/F(x)$ is strictly increasing on $(\tilde{a}^*, \bar{x})$ and strictly decreasing on $(\bar{x}, b^*)$. In both cases, \eqref{XOU1sw} is true if and only if \eqref{XOUcond3} holds.

Alternatively, if \eqref{XOUcond2} doesn't hold, then by in \eqref{XOUhbF1}, the integral $(h_b/F)'$ will always be negative. This means that the function $h_b(x)/F(x)$ is strictly decreasing for all $x \in (-\infty, b^*)$, in which case \eqref{XOU1sw} holds.

We are thus able to show that \eqref{VIJsw} holds, in particular the minimum of 0 is achieved as a result of \eqref{VIJsw1}. To prove \eqref{VIVsw}, we go through a similar procedure. To check that
\begin{align}
(r-\L)\tilde{V}(x) \geq 0
\end{align}
holds, we consider   two cases. First  when $x < b^*$, we get
\begin{align}
(r - \L)\tilde{V}(x)=\frac{e^{b^*}-c_s}{F(b^*)}(r - \L)F(x)=0.
\end{align}
On the other hand, when  $x \geq b^*$, the inequality holds
\begin{align}
(r - \L)\tilde{V}(x)=(r - \L)h_s(x)>0,
\end{align}
since  $b^* > x_s$ (the first inequality of \eqref{XOUb0prop}) and \eqref{Lr}. 

Similarly, when $x \geq b^*$, we have 
\begin{align}
\tilde{V}(x)-(\tilde{J}(x)+h_s(x)) = h_s(x)-h_s(x)=0. 
\end{align}
When $x < b^*$, the inequality holds:
\begin{align}
\tilde{V}(x)-(\tilde{J}(x)+h_s(x)) =\frac{h_s(b^*)}{F(b^*)}F(x) - h_s(x) \geq 0, 
\end{align}
which is equivalent to $\frac{h_s(x)}{F(x)} \leq \frac{h_s(b^*)}{F(b^*)}$,  due to \eqref{eq:XOUhsF1} and \eqref{XOUb0prop}.

\paragraph*{Proof of Theorem \ref{thm:XOU2} (Part 1)} \label{pf:XOU2}
Define the functions
\begin{align}
q_G(x,z)&=\int_x^{+\infty} \Phi(s)(\L-r)h_b(s)ds-\int_z^{+\infty} \Phi(s)(\L-r)h_s(s)ds,\label{qGXOU}\\
q_F(x,z)&=\int_{-\infty}^x \Psi(s)(\L-r)h_b(s)ds - \int_{-\infty}^z\Psi(s)(\L-r)h_s(s)ds.\label{qFXOU}
\end{align}
We look for the points $\tilde{d}^*<\tilde{b}^*$ such that
\begin{align}\label{eq:dbXOU}
q_G(\tilde{d}^*,\tilde{b}^*) =0, \quad \textrm{and}\quad q_F(\tilde{d}^*,\tilde{b}^*) =0.
\end{align}
This is because  these two equations are equivalent to \eqref{eq:solveXOUG} and \eqref{eq:solveXOUF}, respectively.

Now we start to solve the equations by first narrowing down the range for $\tilde{d}^*$ and $\tilde{b}^*$. Observe that
\begin{align}
q_G(x,z)&=\int_x^z \Phi(s)(\L-r)h_b(s)ds + \int_z^\infty\Phi(s)[(\L-r)(h_b(s)-h_s(s)]ds\notag\\
&= \int_x^z \Phi(s)(\L-r)h_b(s)ds - r(c_b + c_s)\int_z^\infty\Phi(s) ds\notag \\&<0,\label{qGnegXOU}
\end{align}
for all $x$ and $z$ such that $x_{b2} \leq x<z$. Therefore, $\tilde{d}^* \in (-\infty,x_{b2})$.

Since $b^* > x_s $ satisfies  $q(b^*)=0$ and $\tilde{a}^* < x_{b2}$ satisfies  \eqref{eq:XOUa}, we have
\begin{align}\label{eq:qFinfXOU}
\lim_{z\to+\infty} q_F(x,z) = \int_{-\infty}^x \Psi(s)(\L-r)h_b(s)ds -q(b^*) -  \int_{b^*}^{+\infty}\Psi(s)(\L-r)h_s(s)ds>0,
\end{align}
for all $x \in (\tilde{a}^*, x_{b2})$.  Also, we note that 
\begin{align}\label{eq:qFdzXOU}
\frac{\partial q_F}{\partial z}(x,z)=-\Psi(z)(\L-r)h_s(z)\begin{cases}
<0 &\, \textrm{ if }\, z<x_s,\\
>0 &\, \textrm{ if }\, z>x_s, 
\end{cases}
\end{align}
 and 
\begin{align}\label{eq:qFxxXOU}
q_F(x,x) = \int_{-\infty}^x \Psi(s)(\L-r)\left[h_b(s) - h_s(s)\right] ds = -r(c_b+c_s) \int_{-\infty}^x \Psi(s)ds < 0.
\end{align}
Then, \eqref{eq:qFdzXOU} and \eqref{eq:qFxxXOU}  imply that there exists a unique function $\beta: [\tilde{a}^*, x_{b2})\mapsto \R$ s.t. $\beta(x) >x_s$ and
\begin{align}\label{eq:XOULx}
q_F(x,\beta(x))=0.
\end{align}
Differentiating \eqref{eq:XOULx} with respect to $x$, we see that 
\begin{align}
{\beta}^\prime\!(x) = \frac{\Psi(x)(\L- r)h_b(x)}{\Psi(\beta(x))(\L - r)h_s(\beta(x))} < 0,
\end{align}
for all $x\in (x_{b1}, x_{b2})$. In addition, by the facts that $b^*>x_s$ satisfies $q(b^*)=0$, $\tilde{a}^*$ satisfies \eqref{eq:XOUa}, and the definition of $q_F$, we have
\begin{align*}
\beta(\tilde{a}^*)=b^*.
\end{align*}

By \eqref{qGnegXOU}, we have $\lim_{x\uparrow x_{b2}}q_G(x,\beta(x))<0$. By computation, we get that  
\begin{align*}
\frac{d}{dx}q_G(x,\beta(x))&= -\frac{\Phi(x)\Psi(\beta(x))-\Phi(\beta(x))\Psi(x)}{\Psi(\beta(x))}(\L - r) h_b(x)\\
&= -\Psi(x)\left[\frac{G(x)}{F(x)}-\frac{G(\beta(x))}{F(\beta(x))}\right] (\L - r) h_b(x)<0,
\end{align*}
for all $x\in (x_{b1}, x_{b2})$. Therefore, there exists a unique $\tilde{d}^*$ such that $q_G(\tilde{d}^*,\beta(\tilde{d}^*))=0$ if and only if
\begin{align}
q_G(\tilde{a}^*,\beta(\tilde{a}^*)) > 0.
\end{align}
The above inequality holds if \eqref{XOU2cond1} holds. Indeed, direct computation yields the equivalence:
\begin{align*}
q_G(\tilde{a}^*,\beta(\tilde{a}^*))&= \int_{\tilde{a}^*}^{+\infty} \Phi(s)(\L-r)h_b(s)ds-\int_{b^*}^{+\infty}\Phi(s)(\L-r)h_s(s)ds\\
&= -\frac{h_b(\tilde{a}^*)}{F(\tilde{a}^*)} - \frac{G(b^*)}{F(b^*)}\int_{-\infty}^{b^*} \Psi(s)(\L-r)h_s(s)ds-\int_{b^*}^{+\infty}\Phi(s)(\L-r)h_s(s)ds\\
&= -\frac{e^{\tilde{a}^*}+c_b}{F(\tilde{a}^*)} + \frac{e^{b^*} - c_s}{F(b^*)}.
\end{align*}
When this solution exists, we have
\begin{align}
\tilde{d}^* \in (x_{b1}, x_{b2}) \textrm{ and } \tilde{b}^* := \beta(\tilde{d}^*) > x_s.
\end{align}

Next, we show that the functions $\tilde{J}$ and $\tilde{V}$ given in \eqref{XOUJsol} and \eqref{XOUVsol} satisfy the pair of VIs in \eqref{VIJsw} and \eqref{VIVsw}. In the same vein as the proof for the Theorem \ref{thm:XOU1}, we  show
\begin{align}
(r - \L)\tilde{J}(x) \geq 0
\end{align}
by examining  the 3 disjoint regions on which $\tilde{J}(x)$ assume different forms. When $x<\tilde{a}^*,$
\begin{align}
(r - \L)\tilde{J}(x) = \tilde{P}(r - \L)F(x)=0.
\end{align}
Next, when $x>\tilde{d}^*,$
\begin{align}
(r - \L)\tilde{J}(x) = \tilde{Q}(r - \L)G(x)=0.
\end{align}
Finally for $x\in [\tilde{a}^*, \tilde{d}^*]$,
\begin{align}
(r - \L)\tilde{J}(x) = (r- \L)(\tilde{K}F(x)-h_b(x)) = -(r - \L)h_b(x)>0,
\end{align}
as a result of \eqref{LrbXOU} since $\tilde{a}^*, \tilde{d}^* \in (x_{b1}, x_{b2})$. 

Next,  we verify that
\begin{align}
(r-\L)\tilde{V}(x) \geq 0.
\end{align}
Indeed, we have $(r-\L)\tilde{V}(x)=\tilde{K}(r-\L)F(x)=0$ for $x<\tilde{b}^*$.  When $x \geq \tilde{b}^*$, we get the inequality $(r-\L)\tilde{V}(x)=(r-\L)(\tilde{Q}G(x)+h_s(x))=(r-\L)h_s(x) >0$ since $\tilde{b}^*>x_s$ and due to  \eqref{Lr}.

It remains to show that  $\tilde{J}(x)-(\tilde{V}(x)-h_b(x))\geq 0$ and $\tilde{V}(x)-(\tilde{J}(x)+h_s(x))\geq 0$.
When $x<\tilde{a}^*$, we have
\begin{align*}
\tilde{J}(x)-(\tilde{V}(x)-h_b(x)) = (\tilde{P}-\tilde{K})F(x) + (e^x+c_b) = -F(x)\frac{e^{\tilde{a}^*}+c_b}{F(\tilde{a}^*)} + (e^x+c_b) \geq 0. 
\end{align*}
This inequality holds since we have shown in the proof of Theorem \ref{thm:XOU1} that $\frac{h_b(x)}{F(x)}$ is strictly decreasing for $x < \tilde{a}^*$.  In addition,
\begin{align*}
\tilde{V}(x)-(\tilde{J}(x)+h_s(x)) = F(x)\frac{e^{\tilde{a}^*}+c_b}{F(\tilde{a}^*)} - (e^{x}-c_s) \geq 0,
\end{align*}
since \eqref{q1XOU} (along with the ensuing explanation) implies that $\frac{h_s(x)}{F(x)}$ is increasing for all $x < \tilde{a}^*$.


In the other region where  $x \in [\tilde{a}^*, \tilde{d}^*]$, we have 
\begin{align*}
&\tilde{J}(x)-(\tilde{V}(x)-h_b(x)) = 0,\\
&\tilde{V}(x)-(\tilde{J}(x)+h_s(x)) = h_b(x)-h_s(x) = c_b +c_s \geq 0.
\end{align*}
When $x> \tilde{b}^*$,  it is clear that 
\begin{align*}
&\tilde{J}(x)-(\tilde{V}(x)-h_b(x)) = h_b(x)-h_s(x) = c_b +c_s \geq 0,\\
&\tilde{V}(x)-(\tilde{J}(x)+h_s(x)) = 0.
\end{align*}

To establish the inequalities for $x\in (\tilde{d}^*,\tilde{b}^*)$,  we first denote
\begin{align*}
g_{\tilde{J}}(x)&:= \tilde{J}(x)-(\tilde{V}(x)-h_b(x)) = \tilde{Q}G(x)-\tilde{K}F(x)+h_b(x) \\
&=F(x)\int_{\tilde{d}^*}^x \Phi(s)(\L-r)h_b(s)ds-G(x)\int_{\tilde{d}^*}^x \Psi(s)(\L-r)h_b(s)ds,\\
g_{\tilde{V}}(x)&:= \tilde{V}(x)-(\tilde{J}(x)+h_s(x)) = \tilde{K}F(x)-\tilde{Q}G(x)-h_s(x)\\
&= F(x) \int_x^{\tilde{b}^*} \Phi(s)(\L-r)h_s(s)ds-G(x)\int_x^{\tilde{b}^*} \Psi(s)(\L-r)h_s(s)ds.
\end{align*}
In turn, we compute to get 
\begin{align*}
g_{\tilde{J}}'(x)&=F'(x)\int_{\tilde{d}^*}^x \Phi(s)(\L-r)h_b(s)ds-G'(x)\int_{\tilde{d}^*}^x \Psi(s)(\L-r)h_b(s)ds,\\
g_{\tilde{V}}'(x)&=F'(x)\int_x^{\tilde{b}^*} \Phi(s)(\L-r)h_s(s)ds-G'(x)\int_x^{\tilde{b}^*} \Psi(s)(\L-r)h_s(s)ds.
\end{align*}
Recall the definition of $x_{b2}$ and $x_s$, and the fact that $G'<0<F'$, we have $g_{\tilde{J}}'(x)>0$ for $x\in (\tilde{d}^*,x_{b2})$ and $g_{\tilde{V}}'(x)<0$ for $x\in (x_s,\tilde{b}^*)$. These, together with the fact that $g_{\tilde{J}}(\tilde{d}^*)=g_{\tilde{V}}(\tilde{b}^*)=0$, imply that
\begin{align*}
g_{\tilde{J}}(x)>0 \textrm{ for }x\in (\tilde{d}^*,x_{b2}), \textrm{ and } g_{\tilde{V}}(x)>0 \textrm{ for } x\in (x_s,\tilde{b}^*).
\end{align*}
Furthermore, since  we have
\begin{align} \label{ineq111}
g_{\tilde{J}}(\tilde{b}^*)=c_b +c_s\geq 0,\quad g_{\tilde{V}}(\tilde{d}^*) =c_b +c_s\geq 0,
\end{align}
and
\begin{align}\label{ineq222}
(\L-r)g_{\tilde{J}}(x)&= (\L-r)h_b(x) <0 \textrm{ for all } x\in (x_{b2},\tilde{b}^*), \\
(\L-r)g_{\tilde{V}}(x)&= -(\L-r)h_s(x) <0 \textrm{ for all } x\in (\tilde{d}^*,x_s).
\end{align}
In view of  inequalities \eqref{ineq111} and \eqref{ineq222},  the maximum principle implies that  $g_{\tilde{J}}(x) \geq 0$ and $g_{\tilde{V}}(x)\geq 0$ for all $x\in (\tilde{d}^*,\tilde{b}^*)$. Hence, we conclude that $\tilde{J}(x)-(\tilde{V}(x)-h_b(x))\geq 0$ and $\tilde{V}(x)-(\tilde{J}(x)+h_s(x))\geq 0$ hold for $x\in (\tilde{d}^*,\tilde{b}^*)$.

\paragraph*{Proof of Theorems \ref{thm:XOU1} and \ref{thm:XOU2} (Part 2)} \label{pf:XOU12}
We now show that the candidate solutions in Theorems \ref{thm:XOU1} and \ref{thm:XOU2}, denoted by $\tilde{j}$ and $\tilde{v}$, are equal to the optimal switching value functions $\tilde{J}$ and $\tilde{V}$ in \eqref{J} and \eqref{V}, respectively. First, we note that $\tilde{j} \leq \tilde{J}$ and $\tilde{v} \leq \tilde{V}$,  since $\tilde{J}$ and $\tilde{V}$ dominate the expected discounted cash low from any admissible strategy. 

Next, we show the reverse inequaities. In Part 1, we have proved that $\tilde{j}$ and $\tilde{v}$ satisfy the VIs \eqref{VIJsw} and \eqref{VIVsw}. In particular, we know that $(r-\L)\tilde{j} \geq 0$, and $(r-\L)\tilde{v} \geq 0$.
Then by Dynkin's formula and Fatou's lemma, as in \citet[p.~226]{Oksendal2003}, for any stopping times $\zeta_1$ and $\zeta_2$ such that $0 \leq \zeta_1 \leq \zeta_2$ almost surely, we have the inequalities
\begin{align}\label{2stoptime}
\E_x\{e^{-r\zeta_1}\tilde{j}(X_{\zeta_1})\} \geq \E_x\{e^{-r\zeta_2}\tilde{j}(X_{\zeta_2})\}, \quad \textrm{and} \quad  \E_x\{e^{-r\zeta_1}\tilde{v}(X_{\zeta_1})\} \geq \E_x\{e^{-r\zeta_2}\tilde{v}(X_{\zeta_2})\}.
\end{align}

For $\Lambda_0=(\nu_1,\tau_1,\nu_2,\tau_2,\dots)$, noting that $\nu_1\leq \tau_1$ almost surely, we have
\begin{align}
\tilde{j}(x) &\geq \E_x\{e^{-r\nu_1}\tilde{j}(X_{\nu_1})\} \label{eq:new1}\\
&\geq \E_x\{e^{-r\nu_1}(\tilde{v}(X_{\nu_1}) - h_b(X_{\nu_1}))\} \label{eq:new2}\\
&\geq \E_x\{e^{-r\tau_1}\tilde{v}(X_{\tau_1})\} - \E_x\{e^{-r\nu_1}h_b(X_{\nu_1})\} \label{eq:new3}\\
&\geq \E_x\{e^{-r\tau_1}(\tilde{j}(X_{\tau_1}) + h_s(X_{\tau_1}))\} - \E_x\{e^{-r\nu_1}h_b(X_{\nu_1})\} \label{eq:new4}\\
&= \E_x\{e^{-r\tau_1}\tilde{j}(X_{\tau_1})\} + \E_x\{e^{-r\tau_1}h_s(X_{\tau_1}) - e^{-r\nu_1}h_b(X_{\nu_1})\},\label{eq:new5}
\end{align}
where \eqref{eq:new1} and \eqref{eq:new3} follow from \eqref{2stoptime}. Also, \eqref{eq:new2} and \eqref{eq:new4} follow from \eqref{VIJsw} and \eqref{VIVsw} respectively. Observing that \eqref{eq:new5} is a recursion and $\tilde{j}(x) \geq 0$ in both Theorems \ref{thm:XOU1} and \ref{thm:XOU2}, we obtain
\begin{align*}
\tilde{j}(x) \geq \E_x\left\{\sum_{n=1}^\infty [e^{-r\tau_n}h_s(X_{\tau_n}) - e^{-r \nu_n} h_b(X_{\nu_n})]  \right\}.
\end{align*} 
Maximizing over all $\Lambda_0$ yields that  $\tilde{j}(x) \geq \tilde{J}(x)$. A similar proof gives $\tilde{v}(x) \geq \tilde{V}(x)$.

\begin{remark} If there is no transaction cost for entry, i.e. $c_b = 0$,  then $f_b$, which is now  a linear function with  a non-zero slope,  has one root $x_0$.  Moreover, we have $f_b(x) > 0$ for $x \in  (-\infty, x_0)$ and $f_b(x)< 0$ for $x \in  (x_0, +\infty)$. This implies that  the entry region must be of the form $(-\infty, d_0)$,  for some number $d_0$. Hence, the  continuation region for entry is the \emph{connected} interval $(d_0, \infty)$. 
\end{remark}

\begin{remark}\label{XOUZervos}  
  Let $\mathcal{L}^\xi$ be the infinitesimal generator of the XOU process $\xi= e^X$, and define the function $H_b(y) := y + c_b \equiv h_b(\ln y)$. In other words, we have the equivalence:
\begin{align}
(\L^\xi-r)H_b(y) \equiv (\L -r)h_b(\ln y).
\end{align}  Referring to \eqref{eqn:XOUfb} and \eqref{Lr}, we have \emph{either} that   
\begin{align}\label{Lxi1}
(\L^\xi-r)H_b(y)\begin{cases}
>0 &\, \textrm{ for }\, y \in (y_{b1}, y_{b2}), \\
<0 &\, \textrm{ for }\, y \in (0, y_{b1}) \cup (y_{b2}, \infty),
\end{cases}
\end{align}
 where $y_{b1} = e^{x_{b1}}>0$ and $y_{b2} = e^{x_{b2}}$ and $x_{b1} < x_{b2}$ are two distinct roots to \eqref{eqn:XOUfb}, \emph{or} 
\begin{align}\label{Lxi2}
(\L^\xi-r)H_b(y) < 0, \quad  \textrm{  for  } ~  y \in (0, y^*) \cup (y^*, \infty),
\end{align}
where $y^* = e^{x^*}$ and $x^*$ is the single root to \eqref{eqn:XOUfb}. In both cases, Assumption 4 of \cite{zervos2011buy}   is violated, and their results cannot be  applied. Indeed,  they would  require that $(\L^\xi-r)H_b(y)$ is strictly negative over  a connected interval of the form $(y_0, \infty)$, for some fixed $y_0\ge 0$.  However, it is clear from \eqref{Lxi1} and \eqref{Lxi2} that such a region is disconnected. 

In fact, the approach by  \cite{zervos2011buy} applies to the optimal switching problems where  the optimal wait-for-entry region (in log-price)   is of the form $(\tilde{d}^*, \infty)$, rather than the \emph{disconnected} region  $(-\infty, \tilde{a}^*) \cup (\tilde{d}^*, \infty)$, as  in our case with an  XOU  underlying. Using the new inferred structure of the wait-for-entry region, we have  modified the arguments in  \cite{zervos2011buy} to solve our optimal switching problem for Theorems \ref{thm:XOU1} and \ref{thm:XOU2}.



\end{remark}

\appendix
\section{Appendix}

\noindent \textbf{A.1 ~Proof of Lemma \ref{lm:HXOU} (Properties of $H$).}\, \label{pf-XOU-H}
The continuity and twice differentiability of $H$ on $(0,+\infty)$ follow directly from those of $h_s$, $G$ and $\psi$. On the other hand,  we have  $H(0):= \lim_{x\to-\infty}\limits \frac{(h_s(x))^+}{G(x)} = \lim_{x\to-\infty}\limits \frac{(e^x-c_s)^+}{G(x)} = \lim_{x\to-\infty} \limits\frac{0}{G(x)} =0$. Hence, the continuity of  $H$   at $0$  follows from
\begin{align*}
\lim_{z\rightarrow 0} H(z) = \lim_{x \to -\infty} \frac{h_s(x)}{G(x)} = \lim_{x \to -\infty} \frac{e^x-c_s}{G(x)} = 0.
\end{align*}
Next, we prove  properties (i)-(iii) of  $H$.

\noindent (i) This  follows trivially from the fact that $\psi(x)$  is a strictly increasing function and $G(x) > 0$.

\noindent (ii) By the definition of $H$,
\begin{align*}
{H}^\prime\!(z) = \frac{1}{\psi'(x)} (\frac{h_s}{G})'(x)= \frac{[e^xG(x)-(e^x-c_s)G'(x)]}{\psi'(x)G^2(x)}, \quad z=\psi(x).
\end{align*}
For $x\in (\ln c_s, +\infty)$, $e^x-c_s >0$,  $G'(x)<0$, so
$e^xG(x)-(e^x-c_s)G'(x)>0$. Also, since both $\psi'(x)$ and $G^2(x)$ are
positive, we  conclude that ${H}^\prime\!(z)>0$ for $z \in (\psi(\ln c_s),
+\infty)$.

The proof of the limit of $H'(z)$ will make use of property (iii), and is thus deferred until after the proof of property (iii).

\noindent (iii) By differentiation, we have
\begin{align*}
{H}^{\prime\prime}\!(z) = \frac{2}{\sigma^2G(x)(\psi'(x))^2}[(\L-r)h_s](x),\quad z=\psi(x).
\end{align*}
Since $\sigma^2, G(x)$ and $(\psi'(x))^2$ are all positive, we only need to determine the sign of $(\L - r)h_s(x) = e^xf_s(x)$. Hence, property (iii) follows from \eqref{Lr}.

To find the limit of $H'(z)$, we first observe that 
\begin{align}\label{hsFlim}
\lim_{x\to +\infty}\frac{h_s(x)}{F(x)} = 0.
\end{align}

Indeed, we have \begin{align*}
\lim_{x \to +\infty}\frac{h_s(x)}{F(x)} &= \lim_{x \to +\infty}\frac{1}{e^{-x}F(x)} = \lim_{x \to +\infty}\left(\int_0^{+\infty} u^{\frac{r}{\mu}-1} e^{(\sqrt{\frac{2\mu}{\sigma^2}}-\frac{1}{u})xu - \sqrt{\frac{2\mu}{\sigma^2}}\theta u -\frac{u^2}{2}} \dx{u}\right)^{-1}\\
&= \lim_{x \to +\infty} \left(\int_0^{\sqrt{\frac{\sigma^2}{2\mu}}} u^{\frac{r}{\mu}-1} e^{(\sqrt{\frac{2\mu}{\sigma^2}}-\frac{1}{u})xu - \sqrt{\frac{2\mu}{\sigma^2}}\theta u -\frac{u^2}{2}} \dx{u} + \int_{\sqrt{\frac{\sigma^2}{2\mu}}}^{+\infty} u^{\frac{r}{\mu}-1} e^{(\sqrt{\frac{2\mu}{\sigma^2}}-\frac{1}{u})xu - \sqrt{\frac{2\mu}{\sigma^2}}\theta u -\frac{u^2}{2}} \dx{u}\right)^{-1}.
\end{align*}
Since the first term on the RHS is non-negative and the second term is strictly increasing and convex in $x$, the limit is zero.

Turning now to $H'(z)$, we note that 
\begin{align*}
{H}^\prime\!(z) = \frac{1}{\psi'(x)} (\frac{h_s}{G})'(x), \quad z=\psi(x).
\end{align*}
As we have shown, for $z> \psi(\ln c_s) \wedge \psi(x_s)$,  $H'(z)$ is a positive and decreasing function. Hence the limit exists and satisfies
\begin{align}\label{H1z}
\lim_{z\to +\infty} H'(z) = \lim_{x\to +\infty} \frac{1}{\psi'(x)} (\frac{h_s}{G})'(x) = c\geq 0.
\end{align} 
Observe that $\lim_{x\to +\infty} \frac{h_s(x)}{G(x)} = +\infty$, $\lim_{x\to +\infty} \psi(x) = +\infty$, and $\lim_{x\to +\infty} \frac{(\frac{h_s(x)}{G(x)})'}{\psi'(x)}$ exists, and $\psi'(x) \neq 0$. We   apply L'Hopital's rule to get
\begin{align}\label{hsFlim2}
\lim_{x\to +\infty} \frac{h_s(x)}{F(x)} = \lim_{x\to +\infty} \frac{\frac{h_s(x)}{G(x)}}{\frac{F(x)}{G(x)}} = \lim_{x\to +\infty} \frac{(\frac{h_s(x)}{G(x)})'}{\psi'(x)}=c.
\end{align}
Comparing \eqref{hsFlim} and \eqref{hsFlim2} implies  that $c=0$. From  \eqref{H1z}, we conclude that    $\lim_{z\to +\infty} H'(z)=0$.
$\scriptstyle{\blacksquare}$\\

\noindent \textbf{A.2 ~Proof of Lemma \ref{lm:hatHXOUr} (Properties of $\hat{H}$).}\, \label{pf-XOU-hatH}
It is straightforward  to check that $V(x)$ is continuous and differentiable
everywhere, and twice differentiable everywhere except at $x=b^{*}$. The same properties hold for  $\hat{h}(x)$. Since both $G$ and $\psi$ are
twice differentiable everywhere, the continuity and differentiability of
$\hat{H}$ on $(0,+\infty)$ and twice differentiability on
$(0,\psi(b^{*})) \cup (\psi(b^{*}),+\infty)$ follow
directly.

To see the continuity of $\hat{H}(y)$ at $0$, note that $V(x) \to 0$ and $e^x \to 0$ as $x \to -\infty$. Then we have \[ \hat{H}(0) := \lim_{x\to-\infty} \frac{(\hat{h}(x))^+}{G(x)} = \lim_{x\to-\infty} \frac{(V(x) -e^x-c_b)^+}{G(x)} = \lim_{x\to-\infty} \frac{0}{G(x)} =0,\] and $\lim_{z \rightarrow 0} \hat{H}(z) = \lim_{x\to-\infty} \frac{\hat{h}(x)}{G(x)}=\lim_{x\to-\infty} \frac{-c_b}{G(x)}=0$. There follows the continuity at $0$.

\noindent (i) For $x \in
    [b^{*},+\infty)$, we have $\hat{h}(x) \equiv -(c_s+c_b)<0$ . Next,  the limits $\lim_{x\to -\infty}\limits V(x) \to 0$ and $\lim_{x\to -\infty}\limits e^x \to 0$ imply that $\lim_{x\to -\infty} \limits \hat{h}(x) = V(x)-e^x-c_b \to -c_b <0$. Therefore, there exists some $\underline{b}$ such that $\hat{h}(x) <0$ for $x \in (-\infty, \underline{b})$. For   the non-trivial case in question, $\hat{h}(x)$ must be positive for some $x$, so we must have $\underline{b} <
b^{*}$.  To conclude, we have  $\hat{h}(x) <0$ for $x \in (-\infty, \underline{b})\cup[ b^{*},+\infty)$. This, along with the facts  that $\psi(x) \in (0,+\infty)$ is a strictly increasing function and $G(x)>0$, implies property (i).

\noindent (ii) By differentiating $\hat{H}(z)$, we get
\begin{align*}
\hat{H}^{'}\!(z) = \frac{1}{\psi'(x)} (\frac{\hat{h}}{G})'(x), \quad z=\psi(x).
\end{align*}
To determine the sign of $\hat{H}^{'}$, we observe that, for $x\geq b^{*}$,
\[(\frac{\hat{h}(x)}{G(x)})' = (\frac{-(c_s+c_b)}{G(x)})'=\frac{(c_s+c_b)G'(x)}{G^2(x)} < 0.\]
Also, $\psi'(x)>0$ for $x\in \R$. Therefore, $\hat{H}(z)$ is
strictly decreasing for $z \geq \psi(b^{*})$.

\noindent (iii) To study the convexity/concavity, we look at the second derivative
\begin{align*}
\hat{H}^{''}\!(z) = \frac{2}{\sigma^2G(x)(\psi'(x))^2}(\L-r)\hat{h}(x),\quad z=\psi(x).
\end{align*}
Since $\sigma^2, G(x)$ and $(\psi'(x))^2$ are all positive, we only need to determine the sign of $(\L - r)\hat{h}(x)$:
\begin{align*}
(\L - r)\hat{h}(x) & = \frac{\sigma^2}{2}({V}^{\prime\prime}\!(x)-e^x)+\mu(\theta-x)({V}^\prime\!(x)-e^x)-r(V(x)-e^x-c_b)\\
&= \begin{cases}
[\mu x - (\mu\theta+\frac{\sigma^2}{2}-r)]e^x+rc_b &\, \textrm{ if }\, x \in (-\infty,b^{*}),\\
r(c_s+c_b) >0 &\, \textrm{ if }\, x\in (b^{*},+\infty).
\end{cases}
\end{align*}
which suggests that $\hat{H}(z)$ is convex for $z\in
(\psi(b^{*}),+\infty)$.


Furthermore,    for $x\in (x_s,b^{*})$, we have
\begin{align*}
(\L - r)\hat{h}(x) &= [\mu x - (\mu\theta+\frac{\sigma^2}{2}-r)]e^x+rc_b= -e^xf_s(x)+r(c_s+c_b)>r(c_s+c_b) >0,
\end{align*}
by the definition of $x_s$.
Therefore, $\hat{H}(z)$ is also convex on
$(\psi(x_s),\psi(b^{*}))$.
Thus far, we have established that $\hat{H}(z)$ is convex on $(\psi(x_s),+\infty)$.

 Next, we  determine the convexity of $\hat{H}(z)$ on $(0,\psi(x_s)]$. Denote $\hat{z}_1 := \argmax_{z\in [0,+\infty)} \hat{H}(z)$. Since  $\sup_{x\in \R}\hat{h}(x) > 0$, we must have
$\hat{H}(\hat{z}_1)=\sup_{z\in [0,+\infty)}\hat{H}(z) > 0$. By its continuity and differentiability,  $\hat{H}$ must be concave at $\hat{z}_1$. Then, there must exist some  interval $(\psi(a^{(0)}),\psi(d^{(0)}))$ over which $\hat{H}$ is concave and $\hat{z}_1 \in (\psi(a^{(0)}),\psi(d^{(0)}))$.

On the other hand, for $x\in (-\infty, x_s]$,
\begin{align*}
((\L - r)\hat{h})'(x) =[\mu x - (\mu\theta+\frac{\sigma^2}{2}-r-\mu)]e^x
\begin{cases}
<0 &\, \textrm{ if }\, x\in (-\infty,x^{*}),\\
>0 &\, \textrm{ if }\, x\in (x^{*}, x_s],
\end{cases}
\end{align*}
where $x^{*}=\theta+\frac{\sigma^2}{2\mu}-\frac{r}{\mu}-1$. Therefore, $(\L - r)\hat{h}(x)$ is strictly decreasing on $(-\infty,x^{*})$,  strictly increasing on $(x^{*}, x_s]$, and is strictly positive at $x_s$ and $-\infty$:
\begin{align*}
(\L - r)\hat{h}(x_s)= r(c_s+c_b)>0 \quad \textrm{and} \quad \lim_{x\to-\infty} (\L - r)\hat{h}(x) = rc_b >0.
\end{align*}
If $(\L - r)\hat{h}(x^{*}) = -\mu e^{x^{*}}+rc_b<0$, then there exist exactly two distinct roots to the equation $(\L - r)\hat{h}(x) =0$, denoted as $x_{b1}$ and $x_{b2}$, such that $-\infty < x_{b1} <x^{*} < x_{b2} < x_s$ and
\begin{align*}
(\L - r)\hat{h}(x) \begin{cases}
>0 &\, \textrm{ if }\, x\in (-\infty,x_{b1})\cup (x_{b2}, x_s],\\
<0 &\, \textrm{ if }\, x\in (x_{b1}, x_{b2}).
\end{cases}
\end{align*}
On the other hand, if $(\L - r)\hat{h}(x^{*}) = -\mu e^{x^{*}}+rc_b\geq 0$, then $(\L - r)\hat{h}(x) \geq 0$ for all $x \in \R$, and $\hat{H} (z)$ is convex for all $z$, which contradicts with the existence of a concave interval. Hence, we conclude that $-\mu e^{x^{*}}+rc_b<0$,  and $(x_{b1}, x_{b2})$ is the unique interval that $(\L - r)\hat{h}(x) <0$. Consequently, $(a^{(0)}, d^{(0)})$ coincides with $(x_{b1}, x_{b2})$ and $\hat{z}_1 \in (\psi(x_{b1}),\psi(x_{b2}))$. This completes the proof. $\scriptstyle{\blacksquare}$\\

\bibliographystyle{apa}
\linespread{-0.3}
\begin{small}
\bibliography{mybib_2012}
\end{small}



\end{document}